\documentclass[12pt]{article}
\usepackage [a4paper,top = 2cm, bottom = 2cm, left = 2cm, right = 2cm]
{geometry}
\usepackage[utf8]{inputenc}
\usepackage{booktabs, multirow}

\usepackage{graphicx}
\usepackage{amsmath, bm}
\usepackage{setspace}
\usepackage{amssymb}
\usepackage{scalerel,stackengine}
\usepackage{hyperref}
\usepackage{geometry}
\usepackage{rotating}
\usepackage{comment}
\usepackage{pdflscape}
\usepackage{fancyhdr}
\usepackage{multirow}
\usepackage{makecell}\usepackage{lscape} 
\usepackage[font=scriptsize]{caption}
\usepackage{subcaption}
\usepackage[]{hyperref}
\hypersetup{
        urlcolor=black,
	colorlinks=true,
	citecolor=blue,
}
\usepackage{amsbsy}
\usepackage[numbers]{natbib}
\setcitestyle{authoryear,open={(},close={)}}
\usepackage{eurosym}
\usepackage{breqn}
\usepackage{fancyhdr}
\usepackage{tabularx}
\usepackage{authblk}
\usepackage{float}
\usepackage{bm}

\usepackage{xurl}

\newcolumntype{C}[1]{>{\raggedright\let\newline\\\arraybackslash\hspace{0pt}}m{#1}}

\title{Heterogeneous drivers of overnight and same-day visits}

\author[1,2]{Francesco Scotti}
\author[1,2]{Andrea Flori}
\author[3]{Piercesare Secchi}
\author[1,2]{Marika Arena}
\author[1,2]{Giovanni Azzone}

\affil[1]{\small\textit{Department of Management, Economics and Industrial Engineering, Politecnico di Milano, Via Lambruschini, 4/B, 20156, Milan, Italy.}}
\affil[2]{\small\textit{
Impact, Department of Management, Economics and Industrial Engineering, Politecnico di Milano}}
\affil[3]{\small\textit{MOX - Dipartimento di Matematica, Politecnico di Milano, Piazza Leonardo da Vinci 32, 20133 Milano, Italy}}

\date{}

\begin{document}

\maketitle

\begin{abstract}



This paper aims to explore the factors stimulating different tourism behaviours, with specific reference to same-day visits and overnight stays. To this aim, we employ mobile network data referred to the area of Lombardy. The paper highlights that larger availability of tourism accommodations, cultural and natural endowments are relevant factors explaining overnight stays. Conversely, temporary entertainment and transportation facilities increase municipalities attractiveness for same-day visits. The results also highlight a trade-off in the capability of municipalities of being attractive in connection to both the tourism behaviours, with higher overnight stays in areas with more limited same-day visits. Mobile data offer a spatial and temporal granularity allowing to detect relevant patterns and support the design of tourism precision policies. 


\vspace{6pt}\textbf{Keywords:} mobile data; tourism; visitors; gravity model; network analysis.

\end{abstract}

\clearpage

\doublespacing

\section{Introduction}

Tourism length of stay constitutes a pivotal concern in tourism demand management since it significantly affects the economic, social and environmental impact of tourism related activities \citep{alegre2006length, barros2010length, alegre2011latent, peypoch2012length, aguilar2019length}. Consistently, the literature on tourism management has formalized the concepts of overnight \textit{tourists} and same-day \textit{visitors} as two alternative tourism behaviours \citep{rodriguez2018length}. More specifically, tourists are delineated as individuals who venture to locations distinct from their usual residence, engaging in overnight stays \citep{rodriguez2018length, nyns2022using}. Conversely, visitors are defined as people experiencing same-day visit at a destination \citep{hall2014tourism, page2020tourism}.

The analysis of dynamics behind tourists and visitors behaviours may support policy makers to design more effective socio-economic development strategies. First, tourists and visitors may have a different economic impact at local level, since the former allow accommodation infrastructures to obtain higher occupation rates, reduce fixed costs, and achieve larger returns \citep{pablo2013tourism, tang2015does, shahzad2017tourism, comerio2019tourism}. Second, the two groups may make a different usage of available infrastructures, with overnight tourists requiring accommodations such as hotels, resorts, or guesthouses, while one-day visitors are more interested in adequate parking facilities and transportation options \citep{adebayo2014optimizing, jovanovia2016infrastructure}. 

Also in terms of destination management, distinct strategies could be put in place. Indeed, overnight tourists may ask for access to a wide set of attractions, amenities and cultural events which necessitate a specific effort in terms of tourism planning and management activities. Conversely, one-day visitors may require well-designed visitor centers and properly managed day-use facilities. Finally, overnight tourists and one-day visitors can have diverse environmental footprints since the different stay duration may imply heterogeneous waste generation, water, energy consumption and CO$_2$ emissions production \citep{sun2014economic, sun2016decomposition, paramati2017effects, castilho2021impacts}. 

Despite the relevance of disentangling such alternative tourism behaviours, extant literature has not deeply investigated the main factors stimulating overnight stays and same-day visits, mainly because traditional tourism data disclosed by national statistical offices do not display temporal and spatial resolution that allow to specifically identify tourists and visitors flows (see e.g., \cite{simini2012universal}, \cite{liu2017understanding}, \cite{pompili2019determinants}, \cite{ giambona2020tourism}, \cite{kim2021tourists} and \cite{ma2022analysis}). As a consequence, available studies analysing the determinants of the length of stay have only considered the flows of people spending at least one night at the destination, neglecting the phenomenon of same-day visits \citep{brida2013factors, alen2014determinant, grigolon2014vacation}.

In order to fill such gap, mobile network data may represent an useful instrument. Indeed, leveraging on information related to the most frequent position of users during days and night hours, they enable to distinguish between tourists and visitors \citep{eurostat2014mobile, onder2016tracing, baggio2018strategic, nyns2022using}.


 Against this background, our paper aims to study the heterogeneity of tourists and visitors movements in Lombardy in 2022, based on mobile network data, in order to comprehend the main drivers of these two types of tourism behaviour. Furthermore, the paper investigates the presence of potential trade-offs in the capability of municipalities to attract at the same extent both types of tourism behaviour. More in detail, we aim to answer to the following research questions:

\textit{RQ1: Which are the main factors explaining the attractiveness of municipalities in terms of tourists and visitors flows?}



\textit{RQ2: Is there a trade-off in the capability of municipalities to attract both tourists and visitors?}

\textit{RQ3: Are spotted patterns stable over the whole calendar year, or do we have evidence of alternative drivers over different seasons?}

Our contributions to the comprehension of the main determinants of tourism behaviour are thus manifold. First, we highlight the relevant characteristics of municipalities fostering tourists or visitors flows. In this way, we fill a relevant gap in extant literature mainly focusing on the drivers of the length of stay of tourists, but overlooking same-day visits \citep{thrane2016students, gomez2019modelling, jackman2020distance}. In particular, we find that one additional accommodation bed contributes to a growth of tourists flows by a percentage between 0.1\% and 0.3\% with respect to visitors volumes. Similarly, the presence of cultural heritage items, ski routes and natural reserves raises the same figure by a portion between 0.1\% and 1.1\%. On the other hand, festivals and transportation facilities such as methane distributors and intermodal nodes increase visitors presences by percentages between 0.1\% and 0.6\% with respect to tourists flows.  
 
 Second, based on a monthly gravity model, we provide evidence that it might be difficult for municipalities to attract at the same time overnight stays and one-day visits, as not necessarily most popular areas in terms of tourists also receive large visitors flows. For instance, municipalities experiencing higher tourists inflows tend to achieve limited visitors levels, highlighting alternative drivers motivating the two types of tourism behaviour. Conversely, nodes with a strategic position in the visitors network, bridging communities of municipalities with limited connections among them (e.g., displaying a high betweenness) result particularly attractive also for overnight stays. We discuss how such municipalities may be of particular interest, since they are able to stimulate high tourists and visitors flows at the same time.

 Finally, we show that mobile phone network data display an adequate level of spatial and frequency granularity, allowing to detect relevant seasonal patterns in terms of tourists and visitors flows, thus representing a valid source of information for policy makers to design  more precise local development strategies \citep{gopalan2010improving}.

The paper is structured as it follows: section \ref{sect: back} summarizes the extant evidence on the main characteristics of origin and destination places attracting people flows and highlights the potential of mobile network data to study alternative types of tourism behaviour. Sections \ref{sect: data} and \ref{sect: met} introduce the data and the empirical methods we employ to investigate our research questions. Section \ref{sect: res} shows the main results and discusses the implications of our analysis. The paper concludes by highlighting the main value-added and contributions of our study to the literature on tourism management.

\section{Theoretical Background}
\label{sect: back}

Our paper relates to two wide and growing streams of literature: on the one hand, research on the determinants of origin-destination places that drive tourism flows; on the other hand, recent research on innovative data sources of tourists flows, such as mobile network data, that may allow to disentangle alternative tourism behaviours at a high frequency and spatially granular scale.

Concerning the first research stream, many studies have investigated the factors driving tourism attractiveness, distinguishing between attributes of origin (demand-side) and destination (offer-side). Such analyses reveal that tourists flows tend to originate in areas that are more densely populated and that are characterized by higher GDP per capita and employment opportunities \citep{zhang2007comparative, massidda2012determinants, ma2022analysis}. Conversely, cultural endowment in terms of museums, art galleries, monuments and archaeological sites, the availability of natural amenities such as parks, coast or mountain areas, as well as leisure attractions and accommodation infrastructures are the main drivers on the tourism supply side \citep{prideaux2005factors, lorenzini2011territorial, pompili2019determinants, giambona2020tourism, simini2021deep, xu2022tourism}. Other factors that may contribute to explaining the attractiveness level of destinations are connected with the quality of public transport, absence of crimes, and the cost of living \citep{cracolici2009attractiveness}.

Recent contributions have also investigated psychological aspects that may affect the intensity of tourists flows in specific locations. They show that uncertainty in vehicle travel time, availability of parking facilities, waiting time for other public means of transport may contribute to increasing tourists psychological fatigue \citep{liu2017understanding, lewis2019understanding, sun2020development, kim2021tourists}. Furthermore, symbolic intangible destination factors may unveil the coolness level of cities, with authentic, rebellious, original and vibrant places being more likely visited by tourists \citep{kock2021makes}.

However, such works are normally based on data aggregated by administrative offices \citep{batista2018analysing, mou2020tourists}, hindering the abstraction of tourists’ profiles at fine spatial and temporal resolution. Another drawback of such conventional data sources is the cost and time consumption of the information collection activities \citep{hu2015extracting, vu2015exploring}, preventing such data from being an adequate input information for fine-grained analyses \citep{jing2020fine}.

The second stream of research is related to the usage of alternative data sources to analyse tourism flows at an adequate spatial and temporal scale, thus solving some limitations of previous studies.  Mobile network data have been proved to be a valid solution to track tourism behaviours. Indeed, compared to the traditional data (e.g. questionnaires and itinerary blogs), mobile network data have a higher level of resolution and reliability \citep{hallo2012gps}. In addition, a comparison of tourists’ recalled diaries with mobile network data showed that the questionnaire-based data varied greatly from the real GPS data, implying there exists a bias in the traditional data that may not be adequately accurate when aiming to explore tourists' behaviors \citep{li2017precision}. Moreover, differently from alternative data sources such as geo-tagged photos and online diaries, that are often unavoidably inconsistent with the actual tourism behavior due to regulatory issues (such as prohibiting photographing, ethics, signal shielding of position sensors, etc.) and usually more widespread among young population cohorts, mobile network data avoid the recording of deviant and redundant information and provide a more representative overview of people flows \citep{salas2018tourists}.

Therefore, mobile network data can be considered as the state-of-the art in terms of quality of information to study tourists flows from a granular geographical and temporal perspective, and such type of information has been widely employed to unveil the spatial structure of macro- and micro-scopic tourists movements \citep{qiu2021spatiotemporal, liu2022cluster}. In particular, based on the application of clustering methods and complex network approaches, such studies have widely documented recurrent patterns in tourists flows, allowing to describe the spatial distribution of tourists, identifying communities and cluster areas characterized by similar intensity of movements \citep{peng2016network, jing2020fine, park2020spatial}.

Nonetheless, such works have currently neglected the social, environmental, economic, demographic, cultural attributes of locations explaining the attractiveness of places \citep{zeng2018pattern, mou2020tourists, kadar2021tourism}. Furthermore, they have not clarified how the detected spatial patterns may vary depending on the type of tourism behaviour, depending on the fact that people just visit a place without the need of an accommodation, or in case they also spend the night in the visited place \citep{zheng2021chinese, liu2022cluster}. 

Disentangling factors that may attract one day visitors and overnight tourists is indeed relevant for policy makers since trips of alternative duration may have a different economic and environmental impact on destinations and may require the design of distinct services and infrastructures. Furthermore, high frequency mobile network data may enable to study how such patterns vary across months due to seasonality. Understanding the distribution of overnight tourists and one-day visitors along the year can aid policymakers in dealing with crowd control and developing appropriate strategies to manage tourists flows effectively. 


Although mobile network data may have a strong potential to explain how local social, economic characteristics and availability of tourism offer and services may attract different levels of tourists and visitors flows, current studies have not employed them to study the main drivers of alternative tourism behaviours.
In this paper, we aim to establish a link between these two streams of literature, showing how mobile network data may be employed to study, at an adequate geographical and temporal level of detail, the main determinants of one day visits and overnight stays. 
In this way, we aim to fill a gap in extant literature that has not documented the drivers of alternative types of tourism and the extent to which such factors may vary along the year depending on seasonality.

\section{Data}
\label{sect: data}

The mobile network data used in this paper have been made available to the authors by Polis, a public entity collaborating with Lombardy region. These data are provided by a main telecommunication company in an anonymised and irreversibly aggregated form, in compliance with the privacy legislation, and the provisions of the EU GDPR, according to the Privacy by Design methodology. These data refer to the calendar year 2022 and cover 163 municipalities in Lombardy as highlighted in Figure \ref{fig: 163 mun} and Table \ref{tab: nomi}. The dataset provides information about the aggregate monthly flow of individuals travelling across each couple of the analysed municipalities with details on the origin and destination of the movement. 

These data allow us to distinguish among two different travelling profiles. In particular, we define as \textit{"visitors"} those individuals who make a visit outside the municipality of usual residence for at least three hours without an overnight stay. On the other hand, we define as \textit{"tourists"} those users with a night cell referring to a municipality that differs from the phone residence. 


Overall, the dataset accounts for 5.4 million tourists presences and 161.5 million visits in Lombardy municipalities in the calendar year 2022.\footnote{We remark that the same individual could be counted more than once as a visitor in the same-day in case the underlying person visits more than one municipality.} Interestingly, Winter is characterized by the largest tourists flows. In particular, December achieves the highest figure with 0.5 million presences. The largest visitors flows are observed in Spring and Autumn, with May and October experiencing the largest values with around 15.5 million people flows. The lowest value is rather observed during August, with 8.8 million flows, probably due to the typical holiday period favouring overstay nights rather than short visits.

 We further describe the tourists and visitors networks, by showing the geographical distribution of the two different types of flows across Lombardy municipalities.

\begin{figure}[ht!]
\centering
\textbf{Top 5\% Tourists Flows}\\
\subfloat[][With Milan]
{\includegraphics[width=.48\textwidth]{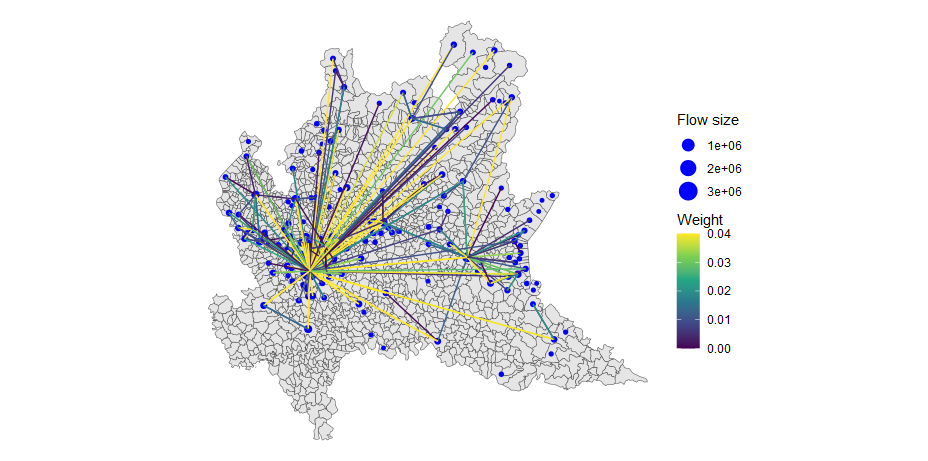}} \quad
\subfloat[][Without Milan]
{\includegraphics[width=.48\textwidth]{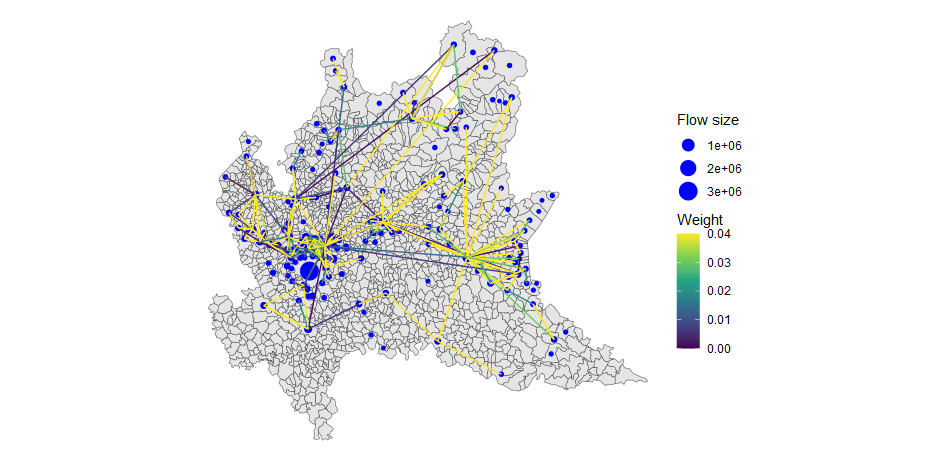}} \\
\textbf{Top 5\% Visitors Flows}
\\
\subfloat[][With Milan]
{\includegraphics[width=.48\textwidth]{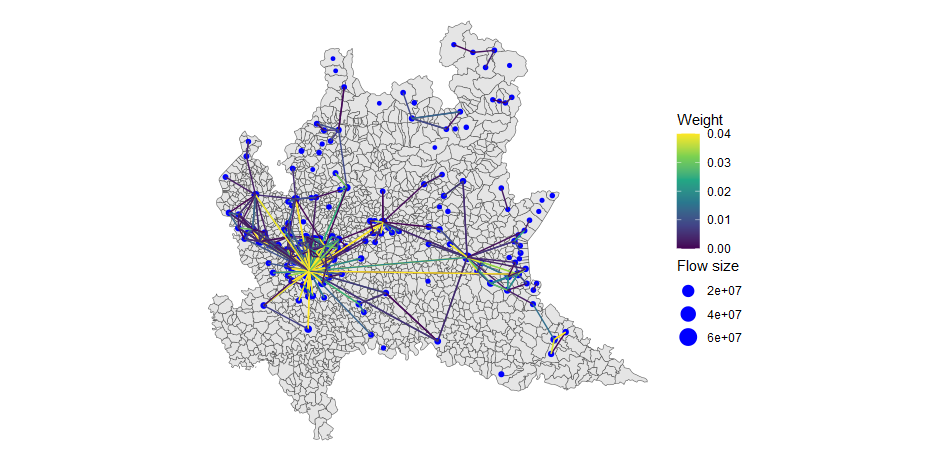}} \quad
\subfloat[][Without Milan]
{\includegraphics[width=.48\textwidth]{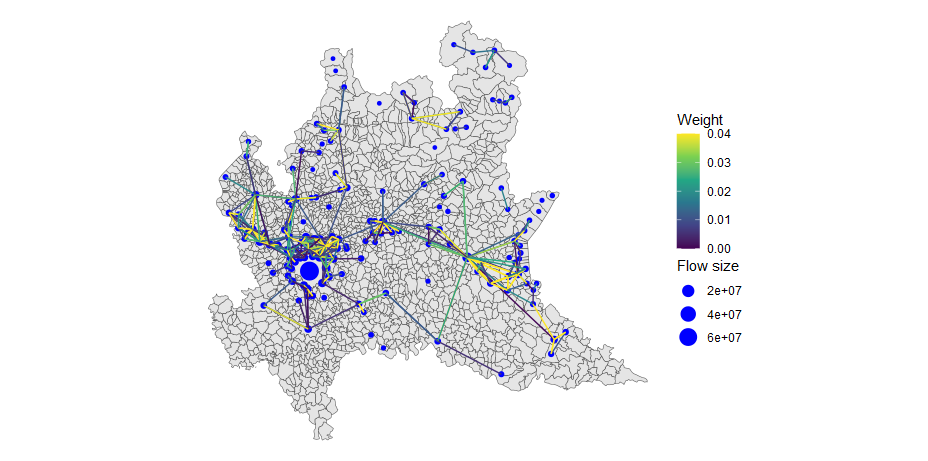}}
\caption{\textbf{Panel top-left: we show the top 5\% flows of tourists across all 163 municipalities. Panel top-right: we show the top 5\% flows of tourists across municipalities excluding Milan. Panel bottom-left: we show the top 5\% flows of visitors across all 163 municipalities. Panel bottom-right: we show the top 5\% flows of visitors across municipalities excluding Milan.}} 
\label{fig: geo flows}
\end{figure}

Figure \ref{fig: geo flows} shows the top 5\% flows across municipalities in the tourists (upper row) and visitors network (lower row) when including (left column) and excluding (right column) the municipality of Milan. We clearly observe that tourists flows tend to connect municipalities characterized by a larger physical distance, whereas visitors flows mainly link neighbour nodes. For instance, when including Milan in the network, the average travel time across the top 5\% flows of tourists is equal to 43.15 minutes (median 30.00), while the same figure accounts for 24.05 minutes (median 20.00) in the visitors network. 

When we exclude Milan, the average travel time in tourists flows decreases to 33.22 minutes (median 24.00), but is still larger than the same figure in the visitors network, where it reaches 22.33 minutes (median 18.00). This result is consistent with the fact that individuals search for closer places when they move for short same-day trips, while they are willing to travel larger distances in case they decide to spend the night in the target location.

\subsection{Data Validation}

Since the dataset is provided by a main telecommunication company, we first check the extent to which they are representative of tourists flows in Italy, by comparing them against information disclosed by the national statistical office (ISTAT).

\begin{figure}[H]
    \centering
    \includegraphics[width=\textwidth]{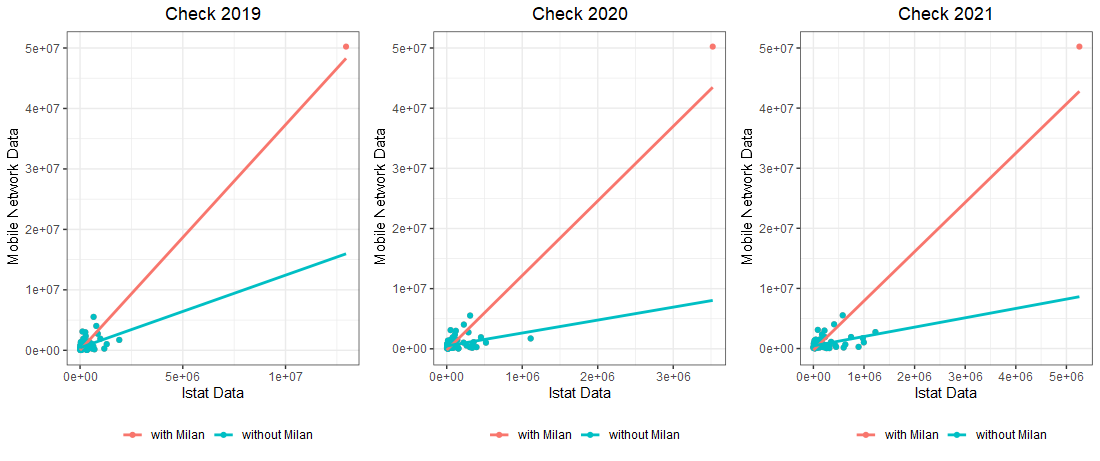}
    \captionsetup{font=scriptsize}
    \caption{\textbf{Correlation between the number of tourists presences in our mobile network and ISTAT data. ISTAT data refer to years 2019, 2020 and 2021. Our mobile network data are related to 2022. In case we consider the whole dataset (including the municipality of Milan) correlation coefficients $\rho$ are equal to 0.97, 0.92 and 0.92 when ISTAT data refer to 2019, 2020 and 2021, respectively (see the light red line). In case we exclude the municipality of Milan the same figures account for 0.61, 0.53 and 0.62, respectively (see the light blue line). In all cases p-values$\sim$ 0. Notice how the correlation coefficient is slightly lower in the years affected by the COVID-19 pandemic (that is not captured by mobile network data referring to 2022).}}
    \label{fig: corr mun}
\end{figure}

As a first step, we match the number of tourists presences in our mobile network data and ISTAT dataset. In particular, we compare our data, available only in 2022, with official statistics provided by ISTAT for the years 2019, 2020 and 2021 (the three most recent years at the moment of writing). Figure \ref{fig: corr mun} highlights the high correlation between ISTAT and our mobile network data with coefficients ranging between 0.92 and 0.97 (p-values $\sim$ 0 in all cases), thus suggesting that we properly capture tourists dynamics (the same figures are 0.61, 0.53 and 0.62 with p-values $\sim$ 0 in all cases, when we exclude Milan from our sample).

\begin{figure}[H]
    \centering
    \includegraphics[width=\textwidth]{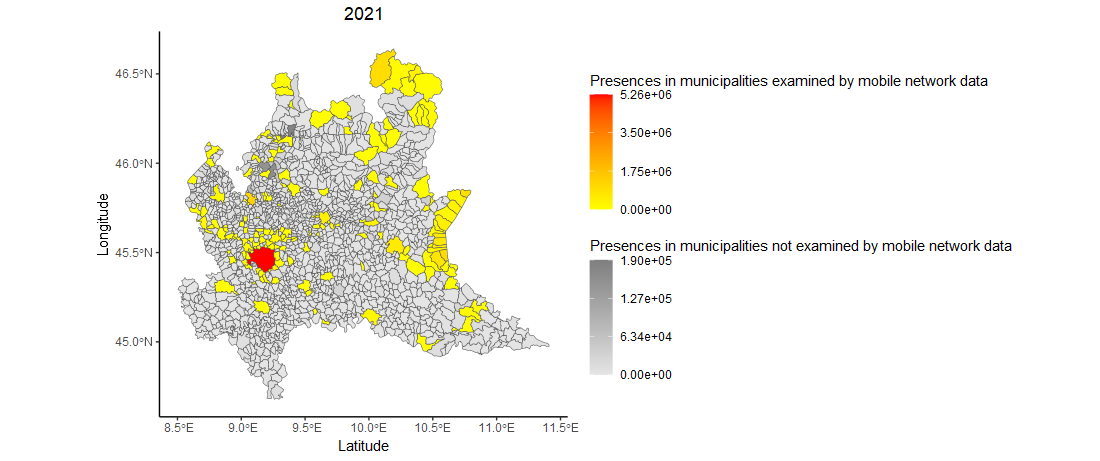}
    \captionsetup{font=scriptsize}
    \caption{\textbf{We show tourist presences in the 163 municipalities included in our analysis in a yellow-red scale of colours. We also show tourist presences in the other municipalities not included (due to the unavailability of our mobile network data) in our analysis in a grey scale of colours. Darker colors refer to larger presence volumes. All data (both those underlying the yellow-red scale and those underlying the grey scale) are disclosed by ISTAT and refer 2021. Similar results hold in 2019 and 2020.}}
    \label{fig: check nc mun pres}
\end{figure}

To further investigate the extent to which our data are representative of tourists flows in Lombardy, we compute the portion of tourists presences covered by the 163 municipalities included in our analysis with respect to total tourists presences in the Lombardy region. To do this, we exploit ISTAT data that cover the entire set of Lombardy municipalities. Figure \ref{fig: check nc mun pres} highlights that the 163 municipalities included in our dataset cover the main tourism areas in Lombardy, accounting for percentages equal to 0.88, 0.86 and 0.85 in 2019, 2020, and 2021 of total tourists presences in Lombardy.\footnote{Such figures are computed as the ratio between the total number of presences of tourists in the 163 municipalities that are included in our analysis (due to the availability of our mobile network data) and the total number of presences of tourists in all Lombardy municipalities. Both the numerator and the denominator are computed based on ISTAT data.}

We finally check the capability of our mobile network data to properly capture seasonality and time patterns along the year in the tourists flows.

We thus compute for each month, the aggregate number of presences observed in our mobile network and ISTAT data with reference to the same set of 163 municipalities included in our analysis. Correlations between the two sets of data are equal to 0.83 (p-value = 0.001), 0.02 (p-value = 0.96), 0.68 (p-value = 0.01) in years 2019, 2020, 2021. It is worthy of noticing that, despite the short available time series, the correlation is significant in year 2019, whereas it is not statistically relevant in year 2020, when tourists presences were disrupted by restrictions against the COVID-19 pandemic.

Correlation exhibits a weak statistical significance also in 2021, highlighting a not yet complete recovery of tourists flows. Indeed, restrictions still affected the winter season in 2021, with limitations related to skiing activities. Such results are consistent with the fact that our mobile network data refer to year 2022, when restrictions to tourists flows were not in place.

Overall, we observe larger values in the mobile network dataset with respect to data disclosed by the national statistical office. This result confirms the evidence provided by \cite{nyns2022using}. In particular, they explain how mobile network data may detect larger tourists flows due to specific factors such as people flows to family/friends, movements towards houses non representing the place of residence,
presences in night owls (nightclubs, parties), leisure events (weddings, festivals, sport competitions),
movements related to the medical sector (hospitals, convalescence) and due to work reasons (e.g., truckers/transporters).

\begin{figure}[H]
    \centering
    \includegraphics[width=\textwidth]{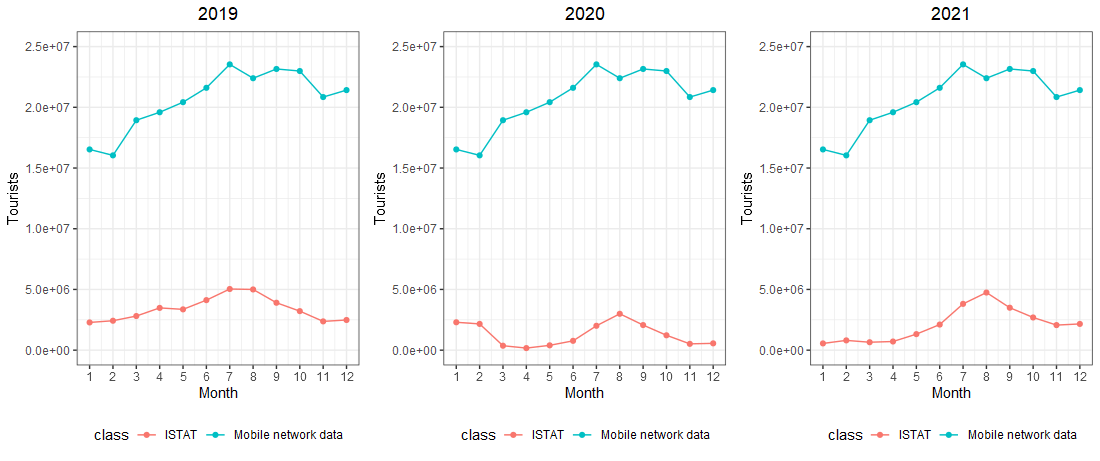}
    \captionsetup{font=scriptsize}
    \caption{\textbf{We show a line plot representing aggregate number of presences observed in our mobile network and ISTAT data with reference to the same set of 163 municipalities included in our analysis. Correlation is equal to 0.83 (p-value = 0.001), 0.02 (p-value = 0.96), 0.68 (p-value = 0.01) in years 2019, 2020, 2021. Mobile network data are always related to year 2022.}}
    \label{fig: ts check}
\end{figure}

\section{Methods}
\label{sect: met}

\begin{figure}[H]
    \centering
    \includegraphics[width=\textwidth]{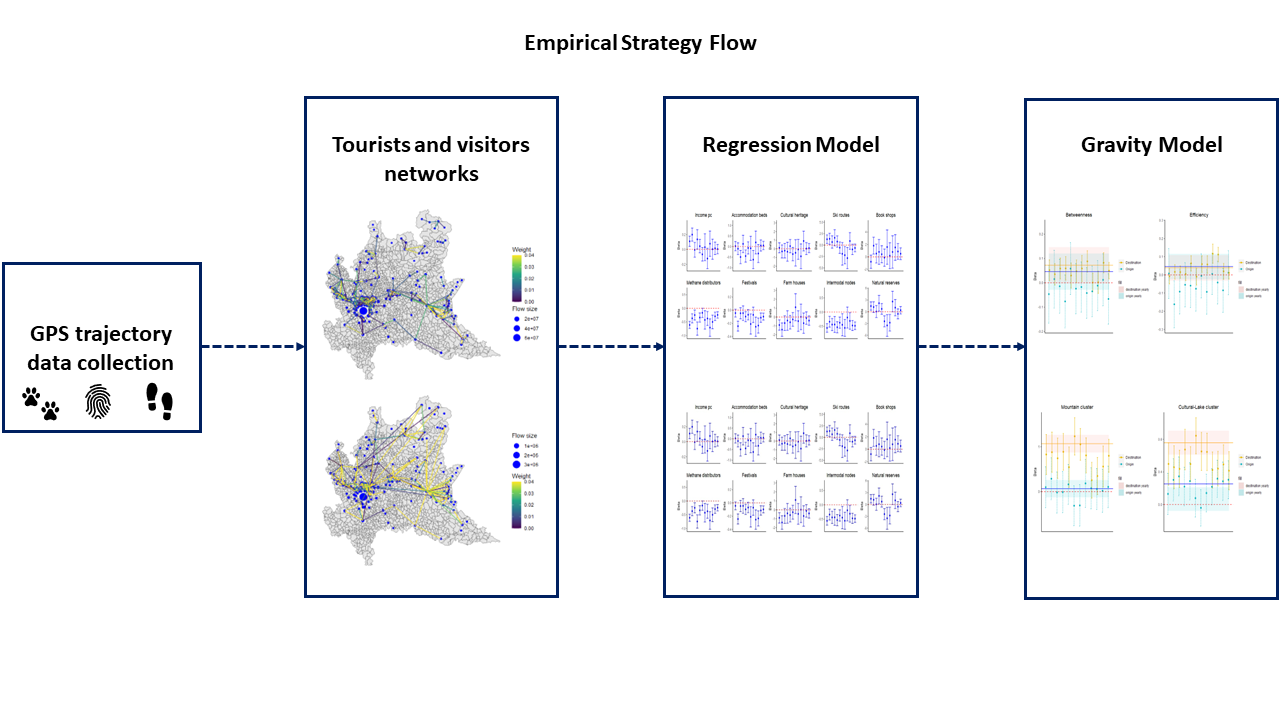}
    \captionsetup{font=scriptsize}
    \caption{\textbf{We show the logical flow of our empirical strategy described in section \ref{sect: met}}.}
    \label{fig: flusso logico}
\end{figure}

In this section, we introduce the empirical strategy that we adopt to address our research questions (see Figure \ref{fig: flusso logico}). 


Section \ref{met: centrality delta} introduces the methodology we apply to answer to our \textit{RQ1}. In particular, we describe the regression models we employ to investigate the drivers of the variation of different centrality measures of municipalities in the tourists with respect to the visitors network. In this way, we clarify the main factors contributing to the attractiveness of municipalities in terms of tourists and visitors flows. 

In addition, section \ref{met: grav} highlights the gravity model used to assess whether centrality indicators in the network of visitors are relevant factors to unveil tourists flows. By the means of this analysis, we aim to discuss our \textit{RQ2}, since we show whether municipalities that receive high tourists flows result to be particularly attractive for visitors, or if instead municipalities tend to receive people with alternative tourism behaviours.

All models described in Sections \ref{met: centrality delta} and \ref{met: grav} are estimated for all months of calendar year 2022. Through the comparison of the results obtained across months, we can thus answer to our \textit{RQ3}, by assessing the stability of identified patterns and the relevance of seasonality factors in explaining tourists flows. 


\subsection{Centrality variation between tourists and visitors networks}
\label{met: centrality delta}

To address our \textit{RQ1}, we analyse the factors mainly fostering tourists and visitors flows. In particular, we investigate the drivers of variation of centrality indicators computed on the network of tourists and visitors. We consider directed weighted networks where municipalities constitute our nodes and the number of people moving between municipalities represent the weight of each edge.

We do this through an OLS model estimated for each month of 2022 to assess the stability of factors influencing tourists and visitors flows along the year and analyse seasonality patterns, as highlighted in our \textit{RQ3}. We rely on the following model:

\begin{equation}
    Y_i = \alpha_0 + \beta *X_i + \epsilon_i
\end{equation}

where $Y_i$ is the percentage variation of the same network indicator computed for municipality $i$ in the network of tourists and visitors.\footnote{More in detail, we compute $Y_i$ with the following formula: 
\begin{equation*}
    Y_i = \frac{(Tourists \ network \ centrality \ indicator - Visitors \ network \ centrality \ indicator)}{Tourists \ network \ centrality \ indicator}
\end{equation*}}

In particular, we focus on instrength, betweenness and efficiency network centrality indicators (see Appendix \ref{app: centrality} for further details on the computation of such variables). We account for such variables since they provide alternative and complementary information on the role of nodes in the networks. The instrength allows us to study the factors that attract the largest volumes of individuals inflows. Through the betweenness, we evaluate the main features of nodes that are frequently part of shortest paths between other nodes in the network, thus potentially representing bridges between communities of nodes with limited connections among them. Finally, the efficiency enables to analyse the extent to which nodes are close to other municipalities, thus having a critical role to foster people flows within the network. 

In terms of regressors, $X_i$ is a vector summarizing the tourism offer in terms of services and attractions provided by each municipality. 

We model as dummy variables all those factors that are not present in the majority of analysed municipalities, where the relevant information is thus related to the availability and not to the number of services or tourism attractions. Such binary variables include \textit{Cultural heritage} items, since the presence of cultural and artistic endowments may increase the tourists flows in a place \citep{massidda2012determinants, giambona2020tourism}. Similarly, natural amenities are considered as factors influencing the attractiveness of locations. For this reason, our model includes the variable \textit{Natural reserves} \citep{lorenzini2011territorial, yang2013modeling}. Furthermore, the presence of leisure and entertainment activities may stimulate additional arrivals in a place. We thus consider the presence of \textit{Ski routes} and \textit{Festivals} \citep{cracolici2009attractiveness, pompili2019determinants}.
As the tourism level is affected by the availability of accommodation infrastructures, we account for the presence of \textit{Farm houses} in each municipality \citep{zhang2007comparative, giambona2020tourism}.
We include the endowments of local transportation infrastructures
by considering whether municipalities present some \textit{Intermodal nodes} and are characterized by the presence of \textit{Methane distributors} since they may affect the likelihood that people transit from such nodes during their trip \citep{lewis2019understanding, sun2020development, kim2021tourists}.

In addition, we consider a set of numerical variables that are related to local economic characteristics or tourism infrastructures and cultural services that are available for the whole sample. In particular, we include the \textit{Income per contribuent} since local economic conditions may foster people outflows or affect the attractiveness level of municipalities \citep{prideaux2005factors, song2010limits, ma2022analysis}. Finally, we account for the number of \textit{Accommodation beds} and the number of \textit{Book shops} per inhabitant.

Descriptive statistics on such regressors and further details on sources of these variables are available in Appendix \ref{app: desc stat centr} in Table \ref{tab: desc stat cent}.

\subsection{Gravity Model}
\label{met: grav}

We further explore the relationship between network indicators computed on the network of visitors and tourists. This analysis contributes to answering to our \textit{RQ2} by explaining whether areas receiving high tourists flows are particularly appealing also for visitors, or if instead municipalities tend to attract alternative tourism behaviours. In particular, we aim to evaluate whether tourists flows can be explained by the centrality of municipalities in the network of visitors according to different centrality indicators. 
Therefore, we employ a gravity model, based on the following equation \citep{van2010gravity}:

\begin{equation}
    Y_{i,j} = G*Z_{i}^{\beta}*Z_{j}^{\gamma}*dist_{i,j}^{\delta}*\epsilon_{i,j}
\end{equation}


The logarithmic version of this model can be estimated through linear models. Therefore, we estimate the following gravity model through a traditional OLS method with a Gaussian error term $\epsilon_{i,j}$.

\begin{equation}
\label{eq: gravity}
    log(Y_{i,j}) = \alpha_0 + \beta*log(Z_{i}) + \gamma*log(Z_{j}) + \delta*log(dist_{i,j}) + \epsilon_{i,j}
\end{equation}

 To also address our \textit{RQ3} by assessing seasonality and the stability of drivers of tourism, we estimate the gravity model described in equation \ref{eq: gravity} for each month of year 2022. In particular, $Y_{i,j}$ is the total aggregate monthly flow of tourists moving from municipality $i$ to municipality $j$. 

 $Z_{i}$ is a vector of characteristics of municipalities of origin of the tourists flow. More in detail, it includes the \textit{Income per contribuent} and the \textit{Population} of the nodes of origin, since a higher wealth and population may foster larger outflows \citep{lorenzini2011territorial, massidda2012determinants, pompili2019determinants, kim2021tourists}. Moreover, it encompasses a set of network indicators describing the centrality of the underlying municipality in the network of visitors. It includes \textit{Instrength}, \textit{Outstrength}, \textit{Betweenness}, \textit{Authority score}, \textit{Hub score}, and \textit{Efficiency} (see Appendix \ref{app: centrality} for more details on how we compute such network centrality indicators). Finally, since people flows may be driven by the local characteristics of territories that attract alternative tourism behaviours, we consider a categorical variable, representing the cluster to which each origin municipality is allocated based on a set of social, economic and environmental variables. Specifically, through this categorical variable we distinguish among municipalities in the \textit{Cultural-Lake}, \textit{Mountain} or \textit{Not Specific} tourism cluster. Details on the cluster analysis are provided in Appendix \ref{app: cluster rob}.

$Z_{j}$ accounts for the same set of factors included in vector $Z_{i}$ but with reference to the municipalities of destination of the tourists flow.

Finally, $dist_{i,j}$ is a vector including alternative measures of distance between node $i$ and $j$. It encompasses the travel distance between the municipality of origin and the municipality of destination.\footnote{The estimation of travel times between all Lombardy’s municipalities is made through r5r, an open-source R package for routing on multi-modal transport networks \citep{pereira2021r5r}. This tool is able to run a simulation model, leveraging geographical data about Lombardy’s municipalities locations and road networks, to obtain the travel times by car.} 

Furthermore, in line with \cite{simini2012universal}, we consider the availability of tourism services and attractions that are within the \textit{Travel distance} between node $i$ and $j$. This is in the spirit of radiation models, suggesting that the distance between areas should be measured not only accounting for geographical distance, but also in terms of density of attractions between the two nodes. In particular, the higher the density of tourism attractions between the two municipalities, the lower the expected flow between such nodes, since people may be attracted by other nodes with a relevant tourism offer between them. 

We consider the total number of \textit{Museums}, \textit{Cultural heritage} items, \textit{Ski routes}, \textit{Farm houses}, \textit{Intermodal nodes}, \textit{Methane distributors} and \textit{Festivals} in the municipalities with a travel distance from the origin lower than that to travel between nodes $i$ and $j$. Each of these variables is multiplied by the average travel distance (from the origin) to reach the nodes in between node $i$ and node $j$ since such “intermediate nodes” may be more attractive and reduce flows between $i$ and $j$ in case they are close and easily accessible from the origin of the movement.

Overall, Appendix \ref{app: grav mod} provides additional details on the set of regressors included in the gravity model introduced by equation \ref{eq: gravity}. In particular, descriptive statistics of network centrality indicators and on the set of variables describing the availability of tourism services and attractions that are within the travel distance between node $i$ and $j$  are included in Table \ref{tab: app inside}.

\section{Empirical Evidence}
\label{sect: res}

\subsection{The drivers of centrality variation in tourists and visitors networks}
\label{met: delta}

In this section, we show the results related to our \textit{RQ1}, through the analysis of the main factors influencing the variation of centrality of municipalities in the tourists network with respect to the visitors network. In particular, we focus on drivers of variation of instrength, betweenness and efficiency over the 12 months of 2022, as described in section \ref{met: centrality delta}. This analysis may thus support policy makers in a better comprehension of the main factors stimulating alternative types of tourism behaviour.

\begin{figure}[H]
    \centering 
     \hspace*{-0.3cm}  
     \textbf{Instrength}
    \begin{subfigure}{\textwidth} 
         \includegraphics[width=\linewidth, height =7.5cm]{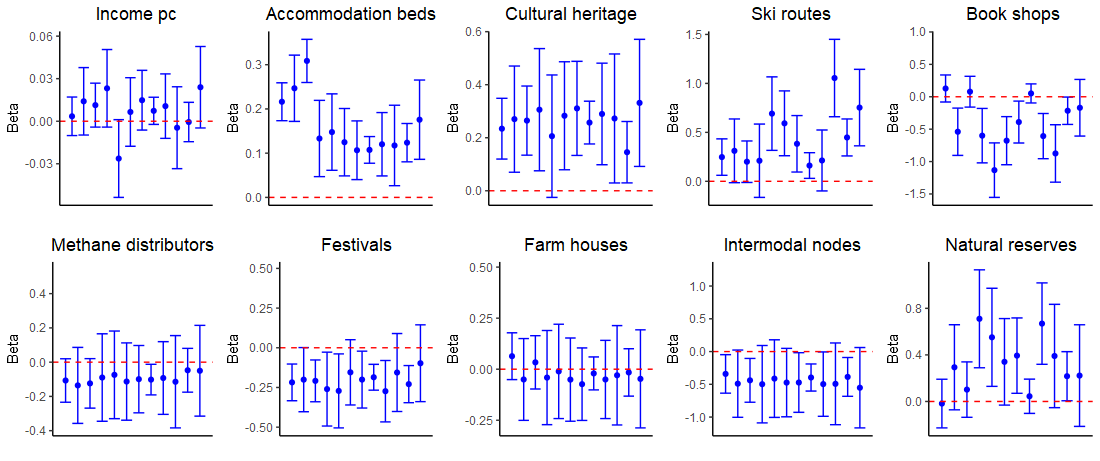}
    \end{subfigure}\hfil 
    \hspace*{0.2cm}
    \textbf{Betweenness}
    \begin{subfigure}{\textwidth}\includegraphics[width=\linewidth, height =7.5cm]{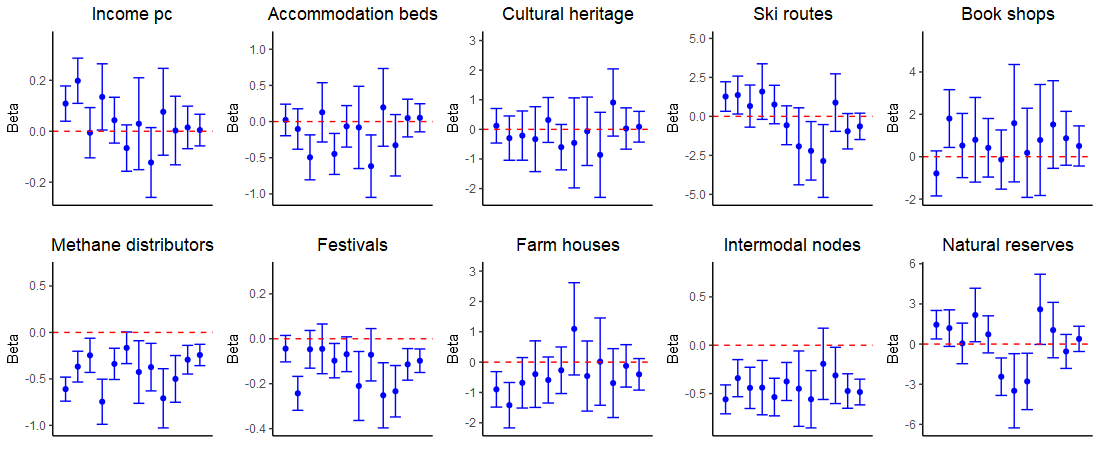}
    \end{subfigure}\hfil 
    \hspace*{0.2cm}
    \textbf{Efficiency}
    \begin{subfigure}{\textwidth}\includegraphics[width=\linewidth, height =7.5cm]{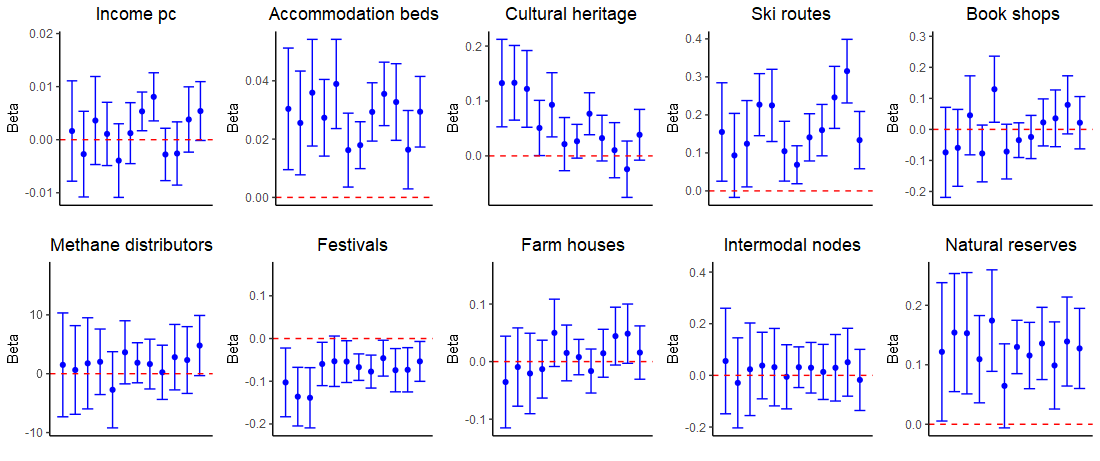}
    \end{subfigure}\hfil 
    \caption{\textbf{We show the monthly coefficients of drivers of delta performances of a set of centrality indicators computed in the tourists and visitors networks. The plots refer to the estimates obtained through the model introduced in section \ref{met: centrality delta}. The upper panel refers to drivers of instrength variation. The middle panel refers to drivers of betweenness variation. The lower panel refers to drivers of efficiency variation.}}
    \label{fig: monthly delta}
\end{figure}

In Figure \ref{fig: monthly delta}, the upper panel shows the coefficients of a set of variables related to tourism services and attractions aiming to explain the variation of instrength in the tourists with respect to the visitors network. Interestingly, one additional accommodation bed induces higher tourists inflows rather than visitors trips  between 0.11\% and 0.31\%. The positive relationship is stable along the year, with a higher magnitude in the first three months of the years. 

Furthermore, the presence of cultural heritage items and ski routes tends to increase the centrality of nodes in the tourists with respect to the visitors network in terms of instrength by a portion between 0.15\% and 0.33\% and between 0.16\% and 1.05\%, respectively. During Spring and Summer months, with the exception of August, also natural reserves contribute to raising tourists with respect to visitors inflows ($\beta$ between -0.01 and 0.71). On the other hand, festivals ($\beta$ between -0.27 and -0.09) and the presence of intermodal nodes ($\beta$ between -0.55 and -0.34) reduce the instrength centrality in the tourist network, as well as the book shops ($\beta$ between -1.13 and 0.12) with the exception of January, March, August and December. 

Such results suggest that the availability of tourism accommodation infrastructures, natural and cultural endowments and services related to ski activities foster overnight stays rather than short visits. On the other hand, temporary entertainment activities (e.g., festivals) or transportation infrastructures, such as intermodal nodes, boost short term visits with respect to tourists inflows.

The middle panel in Figure \ref{fig: monthly delta} analyses the drivers of the percentage variation of betweenness in the tourists with respect to the visitors network. Interestingly, we observe a different pattern with respect to that observed for the instrength. Indeed, in this case accommodation infrastructures, cultural and natural attractions do not foster this centrality indicator in the tourists network. On the other hand, we highlight that for the betweenness, critical factors are represented by transportation services and infrastructures.
Indeed, we identify a negative and statistically significant coefficient along the whole year for methane distributors ($\beta$ between -0.75 and -0.16) and intermodal nodes ($\beta$ between -0.56 and -0.19), thus meaning that higher values of these variables stimulate higher betweenness in the visitors network with respect to the tourists network by percentages between 0.16\% and 0.75\% and between 0.19\% and 0.56\%, respectively.

This result suggests that nodes with higher transportation infrastructures endowments tend to be present in many shortest paths connecting municipalities in the visitors network, meaning that they are key factors to attract short term visits, while they do not represent characteristics stimulating strong tourists flows. We find a similar negative coefficient also for festivals ($\beta$ between -0.27 and -0.05), confirming that such temporary recreational activities may foster short term visits with respect to overnight stays by a percentage between 0.05\% and 0.27\%.

 We observe that the availability of ski routes increases the betweenness in the tourists network during winter months (e.g., January, February), while it raises short term visits in the summer months of August and September, thus highlighting the role of skiing activities in attracting alternative types of tourism behaviours, depending on whether the "snow season" is open or not.

The lower panel of Figure \ref{fig: monthly delta} investigates the factors that contribute to a variation of efficiency in the tourists with respect to the visitors network. 

Similarly to the instrength, we find that a larger availability of accommodation beds ($\beta$ between 0.02 and 0.04), ski routes ($\beta$ between 0.07 and 0.31) and natural reserves ($\beta$ between 0.06 and 0.17) increases this centrality indicator for the tourists network. Also the presence of cultural heritage items ($\beta$ between -0.02 and 0.13) raises the proximity of municipalities to the other nodes in the tourists network, especially in the first half of the year. We rather confirm that festivals ($\beta$ between -0.14 and -0.05) are key drivers of visitors flows. Differently from the instrength case, we do not find a negative impact of intermodal nodes, suggesting that transportation infrastructures do not reduce the centrality of nodes in the tourists with respect to the visitors network. This may be due to the fact that the instrength considers only individuals inflows, whereas the efficiency accounts for both people exiting from and entering into a municipality.

Overall, such results highlight the heterogeneity of drivers of tourists and visitors movements. In particular, this analysis, related to our \textit{RQ1}, shows that the former tend to be more attracted by the availability of accommodation infrastructures, natural and cultural endowments, while transportation services and temporary entertainment activities mainly foster same-day visits.

Moreover, we find that some attractions such as ski routes and natural reserves may further stimulate tourists flows during Spring/Summer and Winter, respectively, thus suggesting the relevance of seasonal patterns, as questioned by our \textit{RQ3}. This evidence may thus allow policy makers to identify the main leverages to design more precise strategies to attract specific types of tourism behaviour.

\subsection{Gravity model evidence}
\label{emp: grav}

Since Lombardy municipalities display relevant percentage variations in terms of instrength, betweenness and efficiency computed on the network of visitors and tourists as highlighted in section \ref{met: delta}, we investigate the extent to which centrality indicators based on same-day visits might be relevant drivers of overnight stays flows. We do this through the gravity model introduced in section \ref{met: grav}. This analysis aims to address our \textit{RQ2} and to study the extent to which areas attracting high levels of tourists are particularly appealing also for visitors, or if instead municipalities tend to attract people with alternative tourism behaviours.

\begin{figure}[H]
\centering
\subfloat[][\emph{}]
{\includegraphics[width=\textwidth, height = 7.5cm]{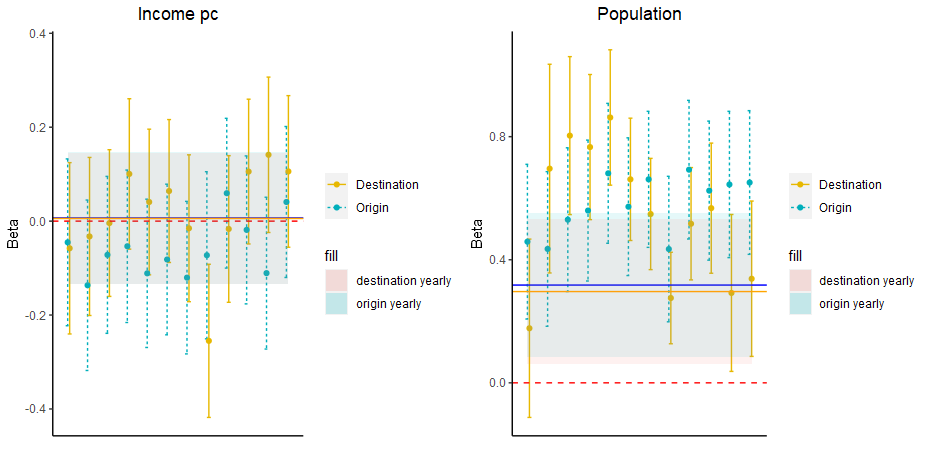}} \\
\subfloat[][\emph{}]
{\includegraphics[width=\textwidth, height = 7.5cm]{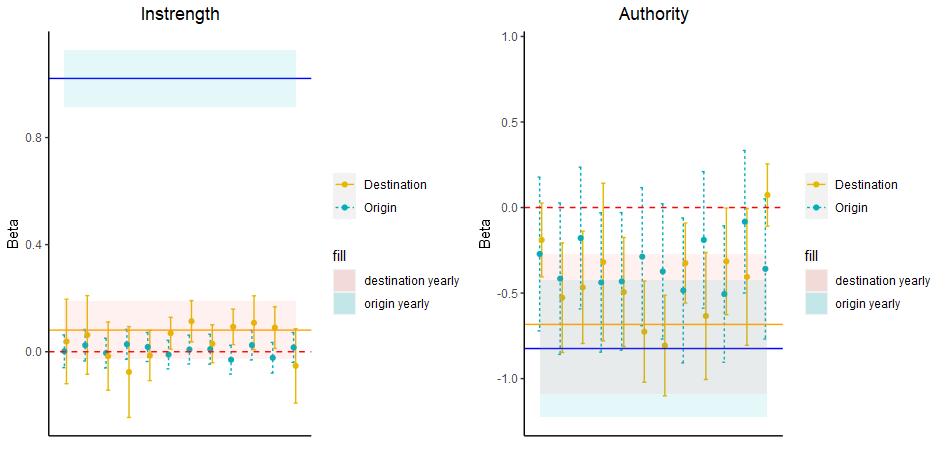}} \\
\subfloat[][\emph{}]
{\includegraphics[width=\textwidth, height = 7.5cm]{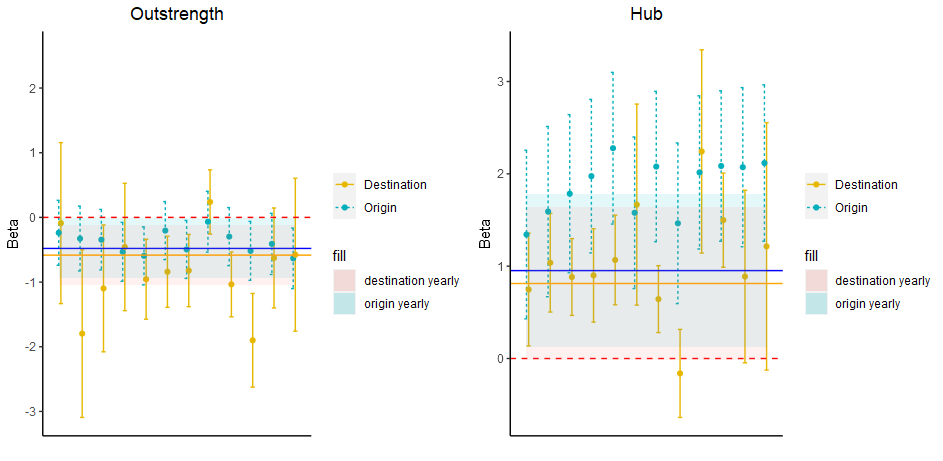}} \\
\caption{\textbf{The monthly coefficients of drivers of monthly flows across municipalities in Lombardy. The plots refer to the estimates obtained through the gravity model introduced in equation \ref{eq: gravity}. Vertical segments refer to the point estimates with 95\% confidence intervals estimated over the 12 months. The horizontal line refers to the point estimates and 95\% confidence intervals of the same gravity model using aggregate yearly (rather than monthly) tourists flows across Lombardy municipalities. Part I.}} 
\label{fig: gravity coeff month 1}
\end{figure}

\begin{figure}[H]
\centering
\subfloat[][\emph{}]
{\includegraphics[width=\textwidth, height = 7.5cm]{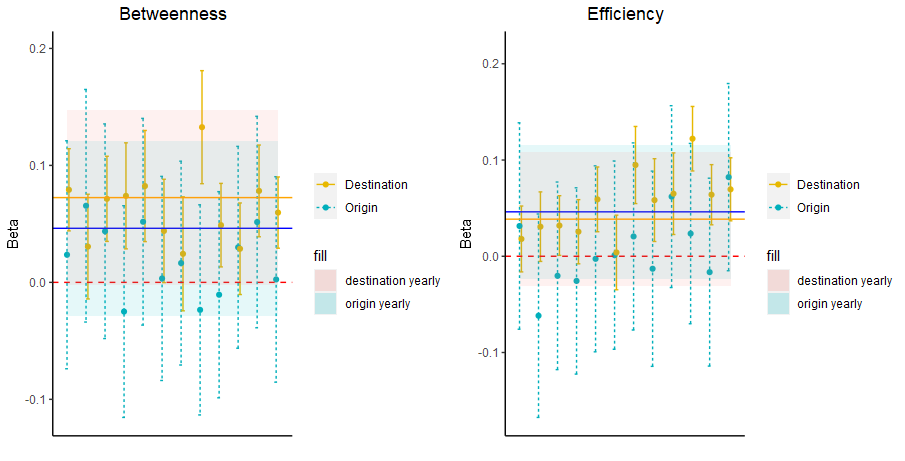}} \\
\subfloat[][\emph{}]
{\includegraphics[width=\textwidth, height = 7.5cm]{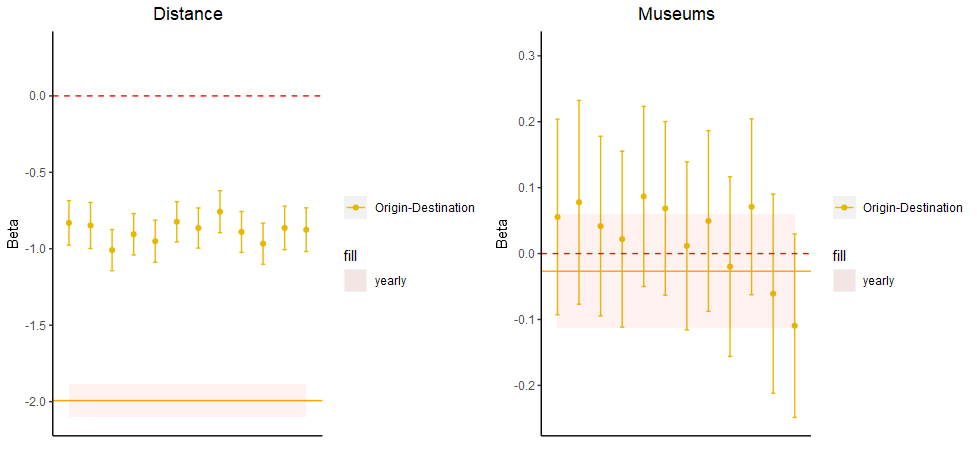}} \\
\subfloat[][\emph{}]
{\includegraphics[width=\textwidth, height = 7.5cm]{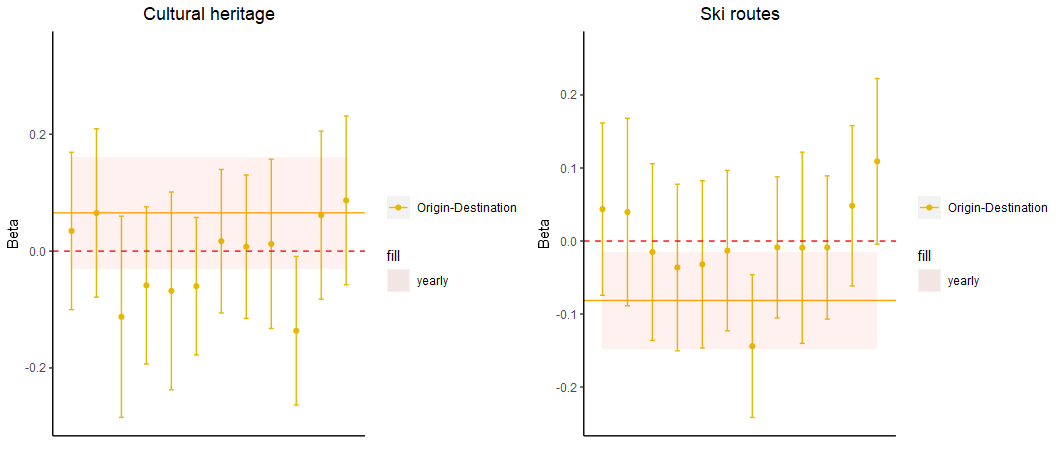}} \\
\caption{\textbf{The monthly coefficients of drivers of monthly flows across municipalities in Lombardy. The plots refer to the estimates obtained through the gravity model introduced in equation \ref{eq: gravity}. Vertical segments refer to the point estimates with 95\% confidence intervals estimated over the 12 months. The horizontal line refers to the point estimates and 95\% confidence intervals of the same gravity model using aggregate yearly (rather than monthly) tourists flows across Lombardy municipalities. Part II.}} 
\label{fig: gravity coeff month 2}
\end{figure}

\begin{figure}[H]
\centering
\subfloat[][\emph{}]
{\includegraphics[width=\textwidth, height = 7.5cm]{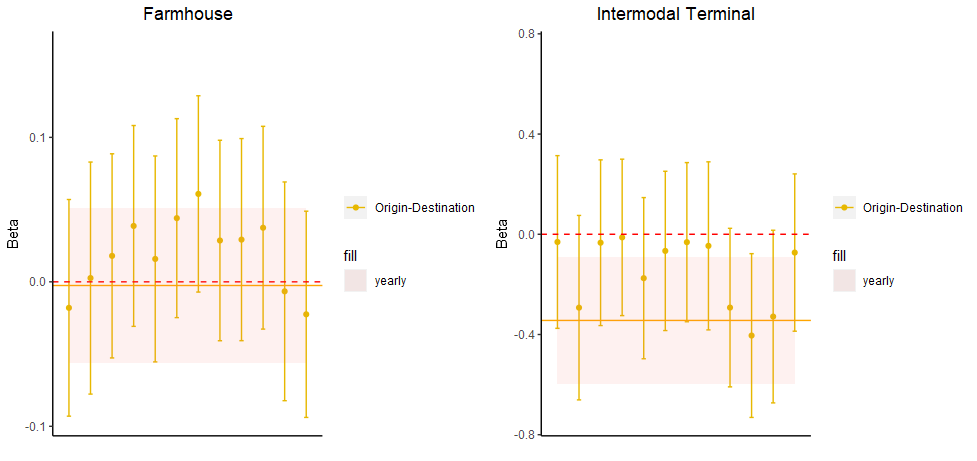}} \\
\subfloat[][\emph{}]
{\includegraphics[width=\textwidth, height = 7.5cm]{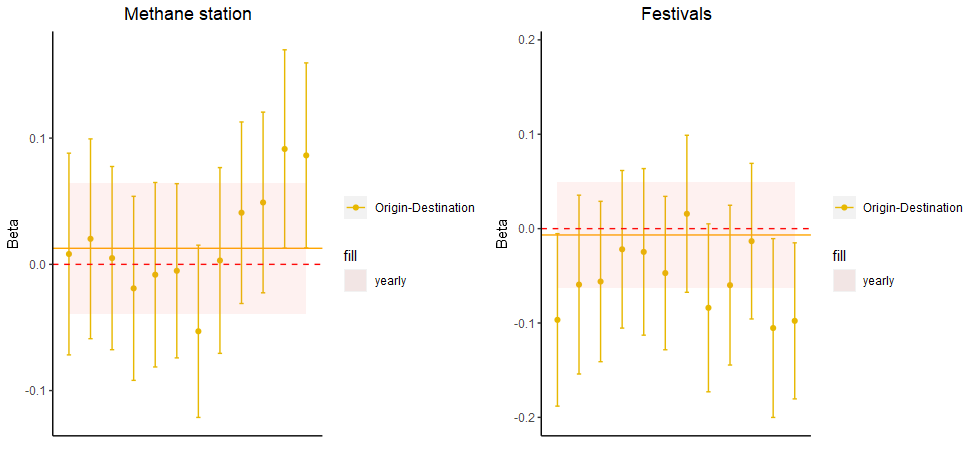}} \\
\subfloat[][\emph{}]
{\includegraphics[width=\textwidth, height = 7.5cm]{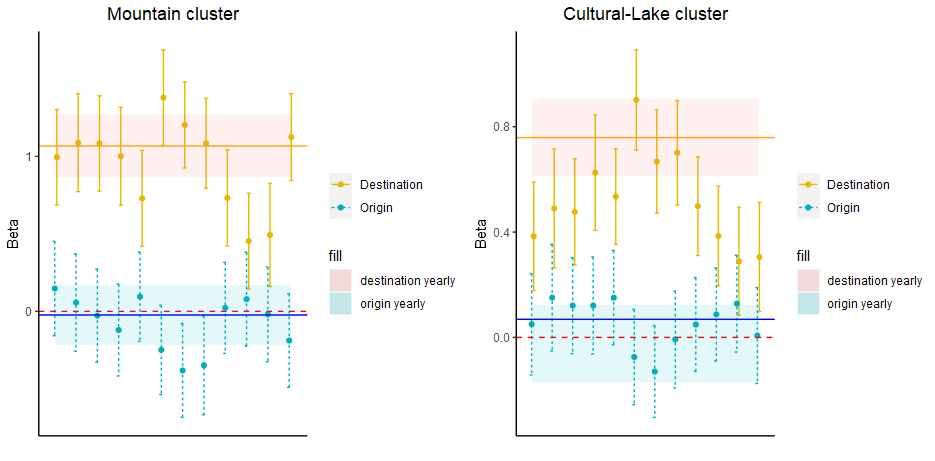}} \\
\caption{\textbf{The monthly coefficients of drivers of monthly flows across municipalities in Lombardy. The plots refer to the estimates obtained through the gravity model introduced in equation \ref{eq: gravity}. Vertical segments refer to the point estimates with 95\% confidence intervals estimated over the 12 months. The horizontal line refers to the point estimates and 95\% confidence intervals of the same gravity model using aggregate yearly (rather than monthly) tourists flows across Lombardy municipalities. Part III.}} 
\label{fig: gravity coeff month 3}
\end{figure}

Figures \ref{fig: gravity coeff month 1}, \ref{fig: gravity coeff month 2} and \ref{fig: gravity coeff month 3} show the estimates of the coefficients of the main drivers of tourists flows across Lombardy municipalities over 12 months of 2022 with related confidence intervals (see Appendix \ref{app: grav mod} and Tables \ref{fig: app coeff month grav 1}, \ref{fig: app coeff month grav 2}, \ref{fig: app coeff month grav 1 (7-12) } and \ref{fig: app coeff month grav 2 (7-12) } for more details). Furthermore, we assess the corresponding coefficient obtained through a model aggregating yearly flows across municipalities. In this way, in line with our \textit{RQ3}, we assess whether coefficients of this set of variables are stable over the year or seasonal patterns strongly affect the main factors stimulating tourism.

Consistently with gravity model theory, we observe the stronger flows of tourists are observed between municipalities with a larger number of inhabitants in line with previous results obtained by \cite{pompili2019determinants} and \cite{giambona2020tourism} for the Italian context. According to the yearly model, a 1\% growth of origin municipality population leads to an increase of tourists flows equal to 0.32\%. The same figure accounts for 0.30\% for destination municipalities. Coefficients with a stronger magnitude and statistically significant are observed when we consider monthly models, with values between 0.44 and 0.69 for origin places and in the range 0.18-0.86 for destinations (the only exception is the month of January where the coefficient is still positive but not statistically significant). 


Interestingly, we find  a not statistically significant estimate for the instrength of destination municipalities over the first half of the year (some exceptions are June, July, September, October and November in the second half of the year). This result suggests that at least over the period January-May, places with large tourists inflows are not necessarily characterized also by a large number of visitors, highlighting the presence of different drivers in play for overnight stays and same-day trips. This result is reinforced by the fact that the coefficient for the yearly model is positive but not statistically significant. On the other hand, more similar determinants between tourists and visitors flows seem to be in place during the Summer and Autumn months.

Such evidence is complemented by the authority score of destinations that tends to exhibit a negative and significant coefficient ($\beta$ between -0.81 and -0.19) with few exceptions (April, December). This is particularly interesting since it can be interpreted as if municipalities with strong tourism inflows are receiving limited visitors flows from hubs in the visitors network. Therefore, the combined evidence of instrength and authority variables directly addresses our \textit{RQ2}, by suggesting that municipalities with large tourists inflows are not necessarily characterized by a large number of same-day visits. Furthermore, such visitors do not come from municipalities that experience large visitors outflows. This evidence also holds at aggregate annual level with a significant $\beta$ = -0.68. Such result corroborates that alternative drivers may motivate tourists and visitors flows, meaning that, for policy makers, it may be complex to attract at the same time alternative types of tourism behaviours.

We observe that the instrength of origin municipalities has a not statistically significant coefficient over the whole year. This means that the number of tourists outflows is not strongly related to the number of visitors inflows. While such relationship is quite stable across months of 2022, we find a different evidence in the aggregate yearly model, where the instrength of origin municipalities has a positive and significant impact on tourists flows ($\beta$ = 1.02). This points to the fact that over the whole year, large tourists outflows are compensated by high visitors inflows. Conversely, we obtain a negative coefficient for the authority of origin places at annual level ($\beta$ = -0.82). Such relationship, that is confirmed at monthly level with a lower statistical significance, suggests that in municipalities with large tourists outflows, visitors inflows do not come from municipality hubs in the visitors network.

We find a different result for the outstrength indicator. For the origin municipalities the coefficient is negative and statistically significant over a significant portion of the year (e.g., in the months of April, May, July, October, November and December with values between -0.63 and -0.41). This pattern shows that places with large outflows of tourists are characterized by a not significant or even negative relationship with the volume of visitors exiting from the same municipality, suggesting that territories tend to experience outflows of individuals with alternative tourism behaviours. The coefficient of the yearly model is negative and statistically significant ($\beta$ = -0.48), meaning that the different patterns of tourists and visitors outflows in the same municipality are stable along the year. 

Also in this case such result can be complemented by the coefficient obtained for the hub score of origin municipalities to address our \textit{RQ2}. The positive and significant ($\beta$ between 1.34 and 2.28) coefficient suggests that territories with large tourists outflows are hubs of the visitors network, meaning that they are municipalities sending visitors towards areas attracting large visitors flows. The combined evidence of outstrength and hub score highlights that places with strong outflows of tourists experience limited visitors outflows. However, those visitors exiting the municipality tend to converge toward areas attracting large short visits inflows. 

Similar evidence holds for places of destination for both outstrength and hub score, thus suggesting that areas attracting larger tourist inflows experience a low number of visitors outflows that tend to be catalyzed by municipalities with large numbers of same-day visits. 

Combining the interpretation of alternative network indicators, we can thus obtain complementary knowledge on the behaviour of tourists and visitors, supporting the identification of more effective tourism management strategies. For instance, being aware of more likely origins and destinations of tourists and visitors flows may allow policy makers to invest on better connections, transportation facilities, customized services and attractions, improving the economic impact of tourism related activities.

We spot a positive interplay over the majority of months of the year for the betweenness of places of destination ($\beta$ between 0.04 and 0.13 and not significant in the months of February, July and October). Such relationship is confirmed at annual aggregate level ($\beta$ = 0.07 and significant with a confidence level of 10\%). This provides evidence that places with large tourists inflows represent nodes belonging to a large number of shortest paths across municipalities in the visitors network. This is particularly relevant since it suggests that nodes bridging alternative communities of areas in the visitors network may be selected also as strategic places to spend the night by tourists. 

We rather find a not statistically significant relationship for the betweenness of places of origin. A consistent result is obtained for the efficiency indicator. Also in this case, we find a positive coefficient for places of destination, with such relationship becoming significant especially during the second half of the year ($\beta$ between 0.06 and 0.12). This result corroborates the previous evidence obtained with the betweenness, confirming that large tourists inflows are experienced by municipalities with a key role in spreading visitors mobility. We do not find a statistically significant relationship for places of origin both at monthly and yearly level. 

Overall, these relationships between tourists flows and the majority of centrality indicators computed in the network of visitors suggest that alternative drivers may have a critical role in stimulating overnight stays and same-day trips, since areas with large tourists flows are not necessarily central in the visitors network. In terms of our \textit{RQ2}, we highlight the presence of a trade-off in the capability of municipalities to attract at the same time large visitors and tourists flows.

In coherence with the gravity model, we also observe a negative and statistically significant coefficient for the travel distance across municipalities ($\beta$ between -1.01 and -0.76), suggesting that stronger tourists flows are observed across municipalities that are closer to each other. Our results confirm previous evidence obtained by \cite{pompili2019determinants} estimating a spatial Durbin model to assess the drivers of international tourism towards Italian provinces. On the other hand, we do not find evidence that a higher density of tourism attractions between municipalities, tends to reduce the flows across such nodes. Few exceptions at a significance level of 10\% are the presence of festivals in January, August, November and December, the presence of ski routes in July and intermodal nodes in September October and November. 

We rather spot interesting patterns for categorical variables related to the cluster assigned to Lombardy municipalities in Appendix \ref{app: cluster rob}, and allowing us to better investigate our \textit{RQ3}.

Our reference cluster is represented by the \textit{Not specific} tourism group. We observe that destinations in the \textit{Mountain} cluster experience larger tourists flows along the year ($\beta$ between 0.45 and 1.38), with coefficient that are higher during Winter months (e.g., December-March) when people have the opportunity to ski or in the Summer (e.g. June, July, August), in a period characterized by milder climate conditions.

Interestingly, also \textit{Cultural-Lake} areas exhibit larger tourists inflows, as suggested by the positive coefficients for destination places. In this case, the coefficient is higher during Summer, whereas it is lower during Winter months, suggesting that tourists prefer spending time in such places in periods characterized by better weather conditions ($\beta$ between 0.29 and 0.90). 
In both cases, we rather do not tend to find evidence of significant coefficients for places of origin, suggesting that drivers to exit the municipality of residence for tourism reasons are homogeneous across clusters over the year.  

Concerning our \textit{RQ3}, the variability of estimated coefficients across different months of the year highlights the need to rely on high frequency data, enabling to account for seasonality patterns to support policy makers taking informed decisions based on robust data driven evidence.

\section{Conclusions}
\label{sect: concl}


In this paper we exploit mobile phone network data to study two alternative types of tourism behaviour: overnight tourists and same-day visitors. We contribute to the extant literature analysing the main factors influencing people length of stay at a destination, whilst the majority of current studies considers only overnight stays, completely neglecting same-day visits.

Our analysis first aims to identify the main determinants of the attractiveness level of Lombardy municipalities, disentangling between tourists and visitors flows. Second, we point to discuss the extent to which places receiving high tourists arrivals result particularly appealing also in terms of visitors. Finally, we highlight whether such patterns are characterized by relevant seasonality or they are stable along the year.

Concerning the first point, we demonstrate that municipalities offering superior tourist services and attractions, including accommodation options, cultural heritage sites, ski routes, and natural reserves, tend to have higher tourists inflows. Conversely, temporary entertainment activities such as festivals and the availability of transportation facilities (e.g., methane stations and intermodal nodes) raise areas attractiveness especially for same-day visits. Such results suggest that policy makers may decide in advance the target tourism behaviour they want to attract in their territory, designing a coherent strategy in terms of services, attractions, and recreational activities.

As regards our second research purpose, the different drivers fostering tourists and visitors flows highlight the presence of a trade-off in the capability of municipalities to attract at the same time overnight stays and one-day visits. We confirm such hypothesis, observing that higher numbers of tourists inflows are exhibited by areas not necessarily receiving large levels of visitors. Similarly, we find that places characterized by large tourists outflows display limited volumes of individuals visiting for a one day trip other places. In this sense, limited exceptions are represented by municipalities with a key role in spreading the mobility in the visitors network, representing nodes that constitute a bridge between communities of municipalities displaying limited connections among them. These places, exhibiting large values of betweenness, account for high flows both in terms of tourists and visitors.

Finally, we demonstrate that mobile phone network data display a high level of temporal and spatial granularity, thus representing an adequate source of information to support the design of tourism management strategies based on real world evidence. We highlight that alternative tourism classes might experience different levels of attractiveness along the year with relevant seasonality patterns. For instance, \textit{Mountain} areas experience larger visits when ski facilities are open or during Summer, probably due to better weather conditions. On the other hand, \textit{Cultural-Lake} territories display higher tourists flows during Spring and Summer, corroborating that climate conditions are also relevant drivers of tourism. Therefore, the time resolution of mobile phone network data can support local policy makers to deal with crowd control and the design of customized tourism services based on the expected volume of people.

Overall, our main findings can be of broad interest for policy makers aiming to design precision policies in the tourism sector, taking into account alternative tourism behaviours.


Despite our effort to implement methodologically grounded research, some limitations still affect this work and may open future research opportunities and discussion. First, our data are disclosed by a relevant telecommunication company but are not provided by national statistical offices. Therefore, rumor and imprecision might be present in our original data, although they have a good match with official statistics. Furthermore, we cover only 163 municipalities in Lombardy. Despite the high representativeness of such tourists flows, future studies may try to include a larger portion of municipalities to assess whether our findings hold on a larger set of territories. Finally, we only focus on the time frame January-December 2022. Next steps of our work might be connected with an extension of the analysed period to evaluate the robustness of our results also on a different calendar year.

\clearpage

\section*{Acknowledgements}

The authors are grateful to Polis, a public entity collaborating with Lombardy region for the design of local policies, for sharing the data on tourism flows in Lombardy during 2022. Furthermore, we extend our acknowledgments also to Vodafone Business and Motion Analytica for their collaboration and crucial expertise in data interpretation, which enriched the outcome of this research.


\section*{Author contributions}
\textbf{Francesco Scotti:} Conceptualization, Methodology, Software, Data curation, Writing- Original draft preparation. \textbf{Andrea Flori:} Conceptualization, Methodology, Writing - Review \& Editing. \textbf{Piercesare Secchi:} Conceptualization, Methodology, Resources, Supervision,  Writing - Review \& Editing.  \textbf{Marika Arena:} Conceptualization, Supervision,  Writing - Review \& Editing. \textbf{Giovanni Azzone:} Conceptualization, Resources, Supervision.

\clearpage

\bibliography{biblio.bib}

\clearpage

\appendix

\setcounter{table}{0}
\setcounter{figure}{0}
\renewcommand{\thetable}{\Alph{section}\arabic{table}}
\renewcommand{\thefigure}{\Alph{section}\arabic{figure}}
\section{Geographical Coverage}
\label{data_app}

\begin{figure}[H]
    \centering
    \includegraphics[width=\textwidth]{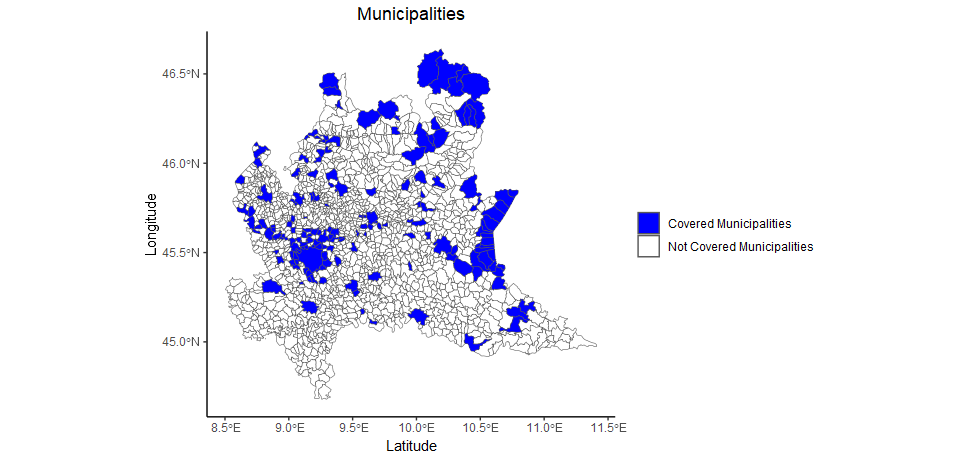}
    \captionsetup{font=scriptsize}
    \caption{\textbf{The geographical distribution of the 163 municipalities covered by our empirical analysis.}}
    \label{fig: 163 mun}
\end{figure}

Figure \ref{fig: 163 mun} shows the geographical distribution of the set of 163 municipalities included in our analysis. Furthermore, Table \ref{tab: nomi} reports the name of the 163 municipalities that are analysed in our paper.

\begin{table}[!htbp] \centering 
  \caption{We list the name of the 163 municipalities that are included in our analysis.} 
  \label{tab: nomi} 
  \scalebox{0.8}{
\begin{tabular}{@{\extracolsep{5pt}} ccc} 
\\[-1.8ex]\hline 
\hline \\[-1.8ex] 
Agrate Brianza & Ferno & Ponte San Pietro \\ 
Albavilla & Gallarate & Ponti Sul Mincio \\ 
Aprica & Garbagnate Milanese & Porlezza \\ 
Arcore & Garbagnate Monastero & Pozzolengo \\ 
Assago & Gardone Riviera & Rezzato \\ 
Bagnatica & Gargnano & Rho \\ 
Bagolino & Grandate & Riva Di Solto \\ 
Baranzate & Grassobbio & Rodengo Saiano \\ 
Basiglio & Gravedona Ed Uniti & Rozzano \\ 
Bergamo & Guardamiglio & Salo \\ 
Binasco & Idro & San Donato Milanese \\ 
Bollate & Iseo & San Felice Del Benaco \\ 
Borgo Virgilio & Lainate & San Giorgio Bigarello \\ 
Bormio & Laveno-Mombello & San Martino In Strada \\ 
Brembate & Lecco & San Pellegrino Terme \\ 
Brescia & Legnano & San Siro \\ 
Bresso & Lezzeno & San Vittore Olona \\ 
Busto Arsizio & Limbiate & Sarnico \\ 
Cambiago & Limone Sul Garda & Saronno \\ 
Campodolcino & Lissone & Segrate \\ 
Cardano Al Campo & Livigno & Sesto Calende \\ 
Carpiano & Lodi & Sesto San Giovanni \\ 
Casalmaggiore & Lomazzo & Settala \\ 
Caspoggio & Lonato Del Garda & Settimo Milanese \\ 
Cassano D'Adda & Lovere & Sirmione \\ 
Castenedolo & Luino & Soiano Del Lago \\ 
Castiglione Delle Stiviere & Maccagno Con Pino E Veddasca & Solbiate Olona \\ 
Castione Della Presolana & Madesimo & Somma Lombardo \\ 
Cavenago Di Brianza & Malgrate & Sondrio \\ 
Centro Valle Intelvi & Mandello Del Lario & Stezzano \\ 
Cernusco Sul Naviglio & Manerba Del Garda & Teglio \\ 
Cerro Maggiore & Mantova & Temu \\ 
Cesano Maderno & Marone & Tignale \\ 
Chiavenna & Menaggio & Tirano \\ 
Chiesa In Valmalenco & Milano & Toscolano-Maderno \\ 
Cinisello Balsamo & Moniga Del Garda & Tremosine Sul Garda \\ 
Clusone & Montano Lucino & Trezzano Sul Naviglio \\ 
Colico & Montichiari & Turate \\ 
Cologno Monzese & Monza & Val Masino \\ 
Como & Monzambano & Valbondione \\ 
Concorezzo & Mozzo & Valdidentro \\ 
Corbetta & Novate Milanese & Valdisotto \\ 
Cornaredo & Novedrate & Valfurva \\ 
Corteno Golgi & Olgiate Olona & Varedo \\ 
Crema & Orio Al Serio & Varese \\ 
Cremona & Orzivecchi & Verdellino \\ 
Cusago & Ospedaletto Lodigiano & Vergiate \\ 
Dalmine & Ossona & Vermiglio \\ 
Darfo Boario Terme & Padenghe Sul Garda & Vezza D'Oglio \\ 
Dervio & Paderno Dugnano & Vigevano \\ 
Desenzano Del Garda & Pavia & Vimercate \\ 
Desio & Pero & Vione \\ 
Dongo & Peschiera Borromeo &  \\ 
Erba & Pieve Emanuele &  \\ 
Erbusco & Ponte Di Legno &  \\ 
\hline \\[-1.8ex] 
\end{tabular} }
\end{table} 

\clearpage

\setcounter{table}{0}
\setcounter{figure}{0}
\renewcommand{\thetable}{\Alph{section}\arabic{table}}
\renewcommand{\thefigure}{\Alph{section}\arabic{figure}}
\section{Centrality Network Indicators}
\label{app: centrality}

In this section we provide details on how we compute the set of network centrality indicators used as drivers of tourists flows in the gravity model introduced in equation \ref{eq: gravity}. In particular:

\begin{itemize}
    \item Instrength: it provides information about the flow entering each node j of the network. In particular, given $w_{i,j}$ the flow of tourists from a node $i$ to a node $j$ of the network, this indicator can be defined as: 

\begin{equation*}
    Instrength_j = \sum_i w_{i,j}
\end{equation*}

\item Outstrength: it provides information about the flow exiting from each node i of the network. In particular, this indicator can be defined as: 

\begin{equation*}
    Outstrength_j = \sum_j w_{i,j}
\end{equation*}

    \item Betweenness: it is the number of shortest paths (connecting all the pairs of nodes of the network) that pass through $i$.

    \begin{equation*}
        Betweenness_i = \frac{\sum_{j,k} n_{j,k}(i)}{n_{j,k}}
    \end{equation*}

    where $n_{j,k}(i)$ is the number of shortest paths connecting node $j$ and $k$ passing through node $i$ and $n_{j,k}$ is the number of shortest paths connecting node $j$ and $k$.

    \item Authority score: it provides information on the centrality of a node based on the sum of hub score centralities of the neighbours (e.g., a node is important if it is pointed by hub nodes).

    \begin{equation*}
        Authority_i = \alpha_1*W^t*Hub-score
    \end{equation*}

    \item Hub score: it provides information on the centrality of a node based on the sum of authority score centralities of the neighbours (e.g., a node is important if it points to authority nodes).

    \begin{equation*}
        Hub score_i = \alpha_2*W^t*Authority score
    \end{equation*}

Therefore, the hub and authority centralities are the eigenvectors of $W^tW$ and $WW^t$ corresponding to
same principal eigenvalue $\lambda = \frac{1}{\alpha_1*\alpha_2}$

    \item Local Efficiency: it provides information about the relevance of a node i in allowing flow within the network.

    \begin{equation*}
      Nodal \  Efficiency_i = \frac{1}{n-1}\sum_{i,j}\frac{1}{d_{i \neq j}}
    \end{equation*}

    \begin{equation*}
        Local \ Efficiency_i = \frac{1}{n}\sum_{i \in G_i} Nodal \  Efficiency(G_i)
    \end{equation*}

    where $G_i$ is the subgraph of neighbours of node $i$.
    
\end{itemize}

\clearpage

\setcounter{table}{0}
\setcounter{figure}{0}
\renewcommand{\thetable}{\Alph{section}\arabic{table}}
\renewcommand{\thefigure}{\Alph{section}\arabic{figure}}
\section{Descriptive statistics drivers of tourists and visitors network centrality variation}
\label{app: desc stat centr}

Table \ref{tab: desc stat cent} shows the descriptive statistics of the drivers of centrality variation in tourists and visitors networks as shown in section \ref{met: delta}.

\begin{table}[!htbp] \centering 
  \scalebox{0.8}{
\begin{tabular}{@{\extracolsep{5pt}} ccccccc} 
\\[-1.8ex]\hline 
\hline \\[-1.8ex] 
 & Q1 & Median & Mean & Q3 & Class  & Source \\ 
\hline \\[-1.8ex] 
Income pc & $17,904.110$ & $20,518.090$ & $20,388.010$ & $22,093.320$ & Numerical & MEF \\ 
Accommodation beds & $0.045$ & $0.128$ & $0.731$ & $0.780$ & Numerical & Lombardy Open Data \\ 
Cultural heritage & $0.000$ & $0.000$ & $0.374$ & $1.000$ & Dummy &  Lombardy Open Data \\ 
Ski routes & $0.000$ & $0.000$ & $0.117$ & $0.000$ & Dummy & Lombardy Open Data \\ 
Book shops & $0.0001$ & $0.0001$ & $0.0002$ & $0.0003$ & Numerical & Lombardy Open Data \\ 
Methane distributors & $0.000$ & $0.000$ & $0.294$ & $1.000$ & Dummy  & Lombardy Open Data \\ 
Festivals & $0.000$ & $0.000$ & $0.313$ & $1.000$  & Dummy & Lombardy Open Data \\ 
Farm houses & $0.000$ & $1.000$ & $0.595$ & $1.000$  & Dummy  & Lombardy Open Data\\ 
Intermodal nodes & $0.000$ & $0.000$ & $0.037$ & $0.000$  & Dummy  & Lombardy Open Data \\ 
Natural reserves & $0.000$ & $0.000$ & $0.067$ & $0.000$  & Dummy  & Lombardy Open Data \\ 
\hline \\[-1.8ex] 
\end{tabular} }
\caption{We highlight the descriptive statistics of the drivers of centrality variation in tourists and visitors networks as shown in section \ref{met: delta}.} 
  \label{tab: desc stat cent} 
\end{table} 

\clearpage

\setcounter{table}{0}
\setcounter{figure}{0}
\renewcommand{\thetable}{\Alph{section}\arabic{table}}
\renewcommand{\thefigure}{\Alph{section}\arabic{figure}}
\section{Cluster analysis approach}
\label{app: cluster rob}

In this section, we explain the empirical approach we use to cluster Lombardy municipalities in different groups based on a set of social, economic and environmental variables, as anticipated in Section \ref{met: grav}. 

In particular, since the local business environment and living conditions are relevant demand side factors influencing the number of tourists flows, we consider the \textit{Income per contribuent}, and the number of \textit{Firms} per inhabitant in the place of origin, consistently with previous studies that use such variables to analyse the main drivers of tourists flows \citep{prideaux2005factors, zhang2007comparative, lorenzini2011territorial, massidda2012determinants, giambona2020tourism, ma2022analysis, xu2022tourism}.

Availability of high quality services may allow to perceive lower fatigue in visiting a specific place \citep{cracolici2009attractiveness, liu2017understanding, lewis2019understanding, sun2020development, kim2021tourists, simini2021deep}. We thus plug in our model the number of \textit{Bank offices} per inhabitant.

As environmental risk may reduce the level of perceived safety and significantly affect the number of visitors in specific periods of the year, we account for \textit{Flood risk} (expressed as the portion of population subject to high flood risk) and \textit{Landslide risk} (expressed as the portion of population subject to high landslide risk) \citep{cracolici2009attractiveness, massidda2012determinants, giambona2020tourism}. We also include the percentage of \textit{Waste sorting}, availability of \textit{Drinking water} (expressed as thousands of cubic meters of water per inhabitant dispensed by the local municipality) and portion of \textit{Soil usage} as proxies of urban ecology, quality of local environmental services and of the level of anthropization of the area \citep{song2010limits, ma2022analysis}.

In terms of demographic variables, we consider \textit{Population density} in line with \cite{lorenzini2011territorial}, \cite{massidda2012determinants} and \cite{pompili2019determinants}.

We finally account for potential heterogeneity across municipalities in the social and healthcare sectors. In particular, we include in our analysis the number of \textit{Schools} per inhabitant, the monetary value of expenditures for \textit{Social services} per inhabitant, the number of \textit{Pharmacies} per inhabitant, and number of beds in \textit{Healthcare infrastructures} per inhabitant.

We perform a cluster analysis based on these social, economic, environmental and demographic variables using the "Ward" hierarchical clustering method as a standard approach to perform such analysis \citep{murtagh2014ward}. We then assess the stability of our results by comparing the output of alternative clustering algorithms, such as the k-means and other hierarchical clustering methods based on alternative agglomeration methods such as "Single", "Complete", "Average", "Mcquitty", "Median", "Centroid" \citep{mcquitty1966similarity, hartigan1975clustering}.

To select the optimal number of clusters we rely on the silhouette coefficient, allowing to compare intra and inter cluster distances, thus providing insight on the quality of the clustering method output \citep{rousseeuw1987silhouettes}. We define the silhouette as:

\begin{equation}
    Silhouette = \frac{1}{N}s_i
\end{equation}

where $s_i$ is the silhouette of observation $i$ and $N$ is the sample size. In particular, $s_i$ can be computed as:

\begin{equation}
    s_i = \frac{b_i -a_i}{max(a_i; b_i)}
\end{equation}

where $a_i$ is the mean distance of observation $i$ from all other units in the same cluster ($c_i$) and $b_i$ is the minimum average distance of observation $i$ from all units in other clusters.

In formula:

\begin{equation}
    a_i = \frac{1}{N_{c_i} -1}\sum_{j \in c_i, j \neq i}(d_{i,j})
\end{equation}

\begin{equation}
    b_i = \operatorname*{min}_{c_l \neq c_i} \frac{1}{N_{c_l}}\sum_{l \in c_l}(d_{i,l})
\end{equation}

where $N_{c_i}$ is the size of cluster $c_i$ and $d_{i,j}$ is the euclidean distance between observation $i$ and $j$.

\begin{figure}[H]
    \centering
    \includegraphics[width=\textwidth]{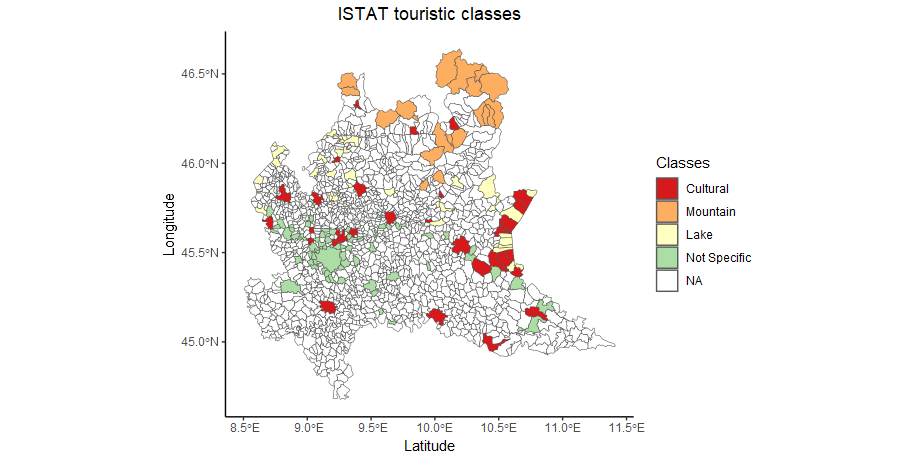}
    \captionsetup{font=scriptsize}
    \caption{\textbf{The tourism classification made by ISTAT of Italian municipalities.}}
    \label{fig: ISTAT cluster}
\end{figure}

Furthermore, we compare the results of our cluster analysis with the tourism classification of Italian municipalities made by ISTAT in 2020.\footnote{Detailed information about the tourism classification of Italian municipalities made by ISTAT in 2020 is available at the following link: \url{https://www.istat.it/it/files//2020/09/classificazione-turistica-comuni.Istat_.pdf.}} Based on ISTAT classification our 163 municipalities are allocated to four alternative classes as reported in Figure \ref{fig: ISTAT cluster}.\footnote{We re-arrange the ISTAT classification, by allocating municipalities with  multiple tourism vocation (e.g. "Cultural-Lake") to the prevalent location. Furthermore, Milan, originally classified by ISTAT in the class "Metropolies" is in this case allocated to the "Not specific - (multidimensional)" tourism class, since no other municipality in Lombardy would have been classified in the same group.} In particular, they belong to the classes \textit{"Cultural"},  \textit{"Mountain"}, \textit{"Lake"}, \textit{"Not specific"}.  

We thus compute the purity index as the percentage of municipalities that are classified by our clustering algorithm in the same tourism class according to the ISTAT analysis. In formula:

\begin{equation}
    Purity \ Index = \frac{1}{N}\sum_{k=1}^{K}max_s|c_k \cap t_s|
\end{equation}

where where $N$ is the number of municipalities, $K$
is the number of clusters, $c_k$ is cluster $k$ and $t_s$ is the classification which has the maximum number of elements in common with cluster $c_k$.

\subsection{Cluster analysis results}

\begin{figure}[H]
    \centering 
     \hspace*{-0.3cm}  
    \begin{subfigure}{\textwidth} 
         \includegraphics[width=\linewidth, height =6cm]{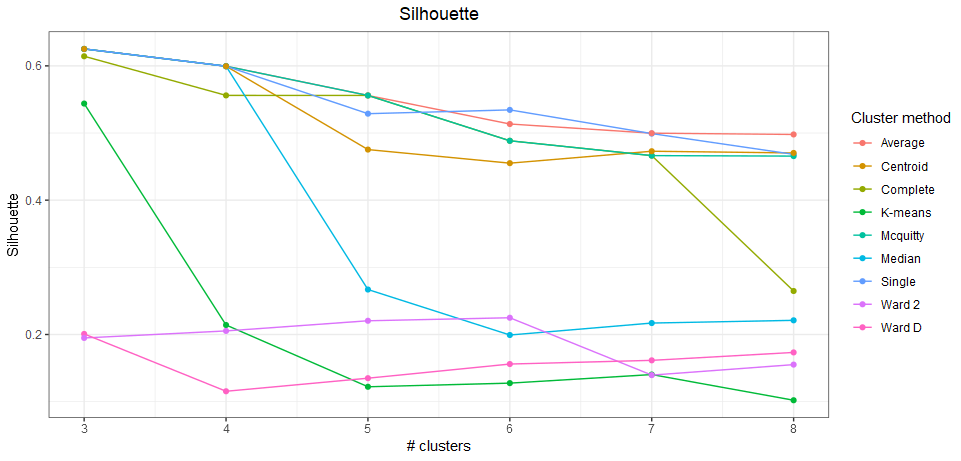}
    \end{subfigure}\hfil 
    \hspace*{0.2cm}
    \begin{subfigure}{\textwidth}\includegraphics[width=\linewidth, height =6cm]{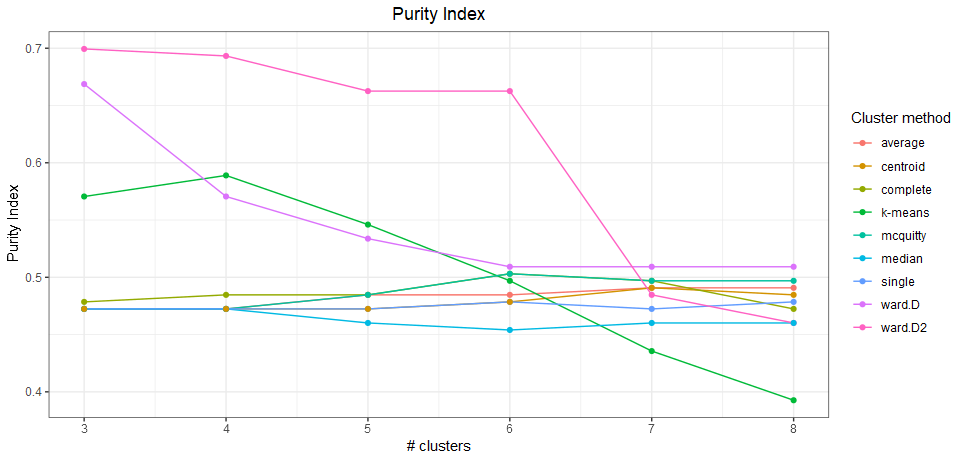}
    \end{subfigure}\hfil 
    \caption{\textbf{The upper panel shows the optimal number of clusters based on the silhouette criterion across alternative hierarchical and non-hierarchical clustering methods. The lower panel shows the purity index for the same set of hierarchical and non-hierarchical clustering methods for a different number of clusters. We use the four tourism classes identified by ISTAT as the reference classification of municipalities to compute the purity index.}}
    \label{fig: numb pur}
\end{figure}

 We first apply the "Ward" agglomeration method \citep{murtagh2014ward} and we select results in correspondence of a number of clusters equal to three since it maximizes the value of the silhouette equal to 0.2 (ranging between 0.12 and 0.17 for a number of clusters between 4 and 8. See the upper panel in Figure \ref{fig: numb pur} for further details). Furthermore, for a number of clusters equal to three we also obtain the maximum purity index accounting for 0.7 (ranging between 0.69 and 0.44 for a number of clusters between 4 and 8). Similar results, suggesting three as the optimal number of clusters, are also confirmed by the set of other clustering algorithms including the k-means and other hierarchical clustering methods based on alternative agglomeration approaches such as "Single", "Complete", "Average", "Mcquitty", "Median", "Centroid" (see the lower panel in Figure \ref{fig: numb pur}).

\begin{figure}[H]
    \centering
    \includegraphics[width=\textwidth]{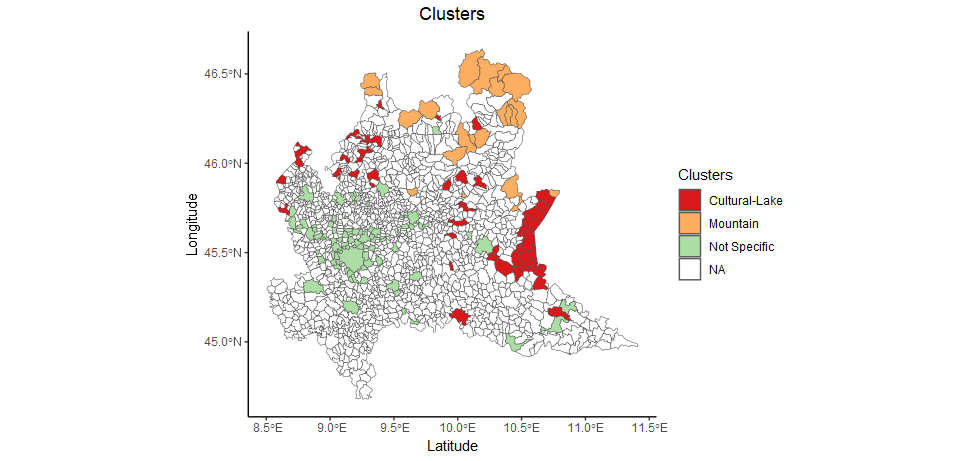}
    \captionsetup{font=scriptsize}
    \caption{\textbf{We show the geographical distribution of Lombardy municipality based on the Ward clustering method \citep{murtagh2014ward}}}
    \label{fig: 3 clust geo}
\end{figure}

Figure \ref{fig: 3 clust geo} shows the geographical distribution of Lombardy municipalities across the three identified clusters. Overall, we notice a certain coherence with the four tourism classes identified by ISTAT and reported in Figure \ref{fig: ISTAT cluster}.  Indeed, the first cluster mainly overlaps with the \textit{Cultural} and \textit{Lake} classes, whereas the second and third clusters mainly represent the \textit{Mountain} and the \textit{Not specific} classes, respectively.

Table \ref{tab: average clust socio econ} shows the average values of the set of social, economic and environmental variables used in the clustering analysis for the three identified groups. We observe that the \textit{Mountain} cluster is characterized by larger landslide and flood risk, while the \textit{Not specific} group accounts for larger income per contribuent, population density, and soil usage. The \textit{Cultural-Lake} cluster exhibits larger social services expenditures and healthcare infrastructures beds.

\begin{table}[H] \centering 
  \caption{We show the average values of a set of social, economic and environmental variables for the three groups identified through our cluster analysis.} 
  \label{tab: average clust socio econ} 
  \scalebox{0.75}{
\begin{tabular}{@{\extracolsep{5pt}} cccc} 
\\[-1.8ex]\hline 
\hline \\[-3.8ex] 
Cluster & $Not-specific$ & $Cultural-Lake$ & $Mountain$ \\ 
\hline \\[-3.8ex] 
Income pc & $22,472$ & $18,472$ & $16,394$ \\ [-0.8ex] 
Soil usage & $0.390$ & $0.116$ & $0.026$ \\ [-0.8ex] 
Waste sorting & $0.740$ & $0.706$ & $0.597$ \\ [-0.8ex] 
Landslide risk & $0.001$ & $0.017$ & $0.083$ \\ [-0.8ex] 
Flood risk & $0.010$ & $0.039$ & $0.058$ \\ [-0.8ex] 
Bank offices & 4.176*10$^{-4}$ & $0.001$ & $0.002$ \\ [-0.8ex] 
Drinking water & $0.105$ & $0.119$ & $0.312$ \\ [-0.8ex] 
Schools & $0.001$ & $0.001$ & $0.002$ \\ [-0.8ex] 
Pharmacies & 3.163*10$^{-4}$ & 4.371*10$^{-4}$ & $0.002$ \\ [-0.8ex] 
Social services & $257.282$ & $366.751$ & $116.130$ \\ [-0.8ex] 
Healthcare infrastructures & 1.353*$^{-5}$ & 4.318*10$^{-5}$ & 2.472*10$^{-4}$ \\ [-0.8ex] 
Population density & $1,944.167$ & $306.023$ & $65.388$ \\ [-0.8ex] 
Firms & $0.076$ & $0.084$ & $0.168$ \\ [-0.8ex] 
\hline \\[-1.8ex] 
\end{tabular} }
\end{table}

\setcounter{table}{0}
\setcounter{figure}{0}
\renewcommand{\thetable}{\Alph{section}\arabic{table}}
\renewcommand{\thefigure}{\Alph{section}\arabic{figure}}
\section{Monthly Gravity Models Results}
\label{app: grav mod}

In this section,  Table \ref{tab: app inside} exhibits the descriptive statistics of a set of variables describing the availability of touristic services and attractions that are within the travel distance between node $i$ and $j$ and used in section \ref{emp: grav} as drivers of tourists flows.

In addition, Tables \ref{fig: app coeff month grav 1}, \ref{fig: app coeff month grav 2}, \ref{fig: app coeff month grav 1 (7-12) } and \ref{fig: app coeff month grav 2 (7-12) } show the coefficients of monthly gravity models estimated in section \ref{emp: grav} and whose coefficients are highlighted in Figures \ref{fig: gravity coeff month 1}, \ref{fig: gravity coeff month 2} and \ref{fig: gravity coeff month 3}.

\begin{table}[!htbp] \centering 
\begin{tabular}{@{\extracolsep{5pt}} ccccc} 
\\[-1.8ex]\hline 
\hline \\[-1.8ex] 
 & Q1 & Median & Mean & Q3 \\ 
\hline \\[-1.8ex] 
Cultural heritage inside & $26.333$ & $355.500$ & $1,144.830$ & $1,859.016$ \\ 
Ski routes inside & $0.000$ & $0.000$ & $185.202$ & $140.669$ \\ 
Farm houses inside & $50.333$ & $620.000$ & $1,837.961$ & $2,939.726$ \\ 
Intermodal nodes inside & $0.000$ & $58.474$ & $144.253$ & $281.339$ \\ 
Methane distribuitors inside & $32.500$ & $489.431$ & $1,143.478$ & $2,050.331$ \\ 
Festivals inside & $29.250$ & $366.583$ & $967.408$ & $1,536.364$ \\ 
Museums inside & $29.333$ & $288.000$ & $931.598$ & $1,517.520$ \\ 
Travel distance & $31.000$ & $53.000$ & $69.282$ & $94.757$ \\ 
\hline \\[-1.8ex] 
\end{tabular} 
 \caption{We show the descriptive statistics of a set of variables describing the availability of touristic services and attractions that are within the travel distance between node $i$ and $j$ and used in section \ref{emp: grav} as drivers of tourists flows. We compute such variables as the product between the average travel distance (from the origin) to reach nodes in between node $i$ and node $j$ and  total number of \textit{Museums}, \textit{Cultural heritage} items, \textit{Ski routes}, \textit{Farm houses}, \textit{Intermodal nodes}, \textit{Methane distributors} and \textit{Festivals} in the municipalities with a travel distance from the origin lower than that to travel between nodes $i$ and $j$.} 
  \label{tab: app inside} 
\end{table}

\begin{table}[!htbp] \centering 
  \scalebox{0.80}{
\begin{tabular}{@{\extracolsep{5pt}}lcccccc} 
\\[-1.8ex]\hline 
\hline \\[-1.8ex] 
 & \multicolumn{6}{c}{\textit{Dependent variable:}} \\ 
\cline{2-7} 
\\[-1.8ex] & \multicolumn{6}{c}{Tourists flows} \\ 
\\[-1.8ex] & (1) & (2) & (3) & (4) & (5) & (6)\\ 
\hline \\[-1.8ex] 
 Income pc orig. & $-$0.045 & $-$0.137 & $-$0.072 & $-$0.054 & $-$0.111 & $-$0.082 \\ 
  & (0.091) & (0.093) & (0.085) & (0.083) & (0.081) & (0.082) \\ 
  & & & & & & \\ [-0.6em]
 Population orig. & 0.459$^{***}$ & 0.435$^{***}$ & 0.530$^{***}$ & 0.560$^{***}$ & 0.681$^{***}$ & 0.572$^{***}$ \\ 
  & (0.128) & (0.128) & (0.119) & (0.117) & (0.116) & (0.114) \\ 
  & & & & & & \\ [-0.6em]
 Instrength orig. & 0.002 & 0.024 & $-$0.005 & 0.028 & 0.018 & $-$0.011 \\ 
  & (0.031) & (0.030) & (0.028) & (0.029) & (0.028) & (0.027) \\ 
  & & & & & & \\ [-0.6em]
 Outstrength orig. & $-$0.238 & $-$0.327 & $-$0.344 & $-$0.532$^{**}$ & $-$0.594$^{***}$ & $-$0.204 \\ 
  & (0.256) & (0.256) & (0.239) & (0.232) & (0.229) & (0.229) \\ 
  & & & & & & \\ [-0.6em]
 Betweenness orig. & 0.024 & 0.066 & 0.044 & $-$0.025 & 0.052 & 0.003 \\ 
  & (0.050) & (0.051) & (0.047) & (0.046) & (0.045) & (0.045) \\ 
  & & & & & & \\ [-0.6em]
 Authority orig. & $-$0.272 & $-$0.416$^{*}$ & $-$0.178 & $-$0.438$^{**}$ & $-$0.432$^{**}$ & $-$0.287 \\ 
  & (0.229) & (0.226) & (0.211) & (0.208) & (0.205) & (0.206) \\ 
  & & & & & & \\ [-0.6em]
 Hub orig. & 1.342$^{***}$ & 1.591$^{***}$ & 1.785$^{***}$ & 1.975$^{***}$ & 2.277$^{***}$ & 1.578$^{***}$ \\ 
  & (0.466) & (0.470) & (0.437) & (0.424) & (0.419) & (0.419) \\ 
  & & & & & & \\ [-0.6em]
 Efficiency orig. & 0.031 &  $-$0.062 &  $-$0.020 &  $-$0.026 &  $-$0.003 & 0.001 \\ 
  & (0.055) & (0.054) & (0.050) & (0.049) & (0.049) & (0.050) \\ 
  & & & & & & \\ [-0.6em]
 Mount clst. orig. & 0.147 & 0.057 & $-$0.028 & $-$0.121 & 0.095 & $-$0.249$^{*}$ \\ 
  & (0.155) & (0.160) & (0.153) & (0.151) & (0.147) & (0.147) \\ 
  & & & & & & \\ [-0.6em]
 Cultural-Lake clst. orig. & 0.050 & 0.151 & 0.121 & 0.121 & 0.151$^{*}$ & $-$0.074 \\ 
  & (0.098) & (0.103) & (0.093) & (0.094) & (0.091) & (0.092) \\ 
  & & & & & & \\ [-0.6em]
\hline 
\hline \\[-1.8ex] 
\textit{Note:}  & \multicolumn{6}{r}{$^{*}$p$<$0.1; $^{**}$p$<$0.05; $^{***}$p$<$0.01} \\ 
\end{tabular}}
\caption{We show the coefficients of drivers of the gravity model highlighted in Figures \ref{fig: gravity coeff month 1}, \ref{fig: gravity coeff month 2}, \ref{fig: gravity coeff month 3}. Column 1-6 refers to models estimated for the first six months of the year 2022. Part I.} 
  \label{fig: app coeff month grav 1} 
\end{table}

\begin{table}[!htbp] \centering 
  \scalebox{0.80}{
\begin{tabular}{@{\extracolsep{5pt}}lcccccc} 
\\[-1.8ex]\hline 
\hline \\[-1.8ex] 
 & \multicolumn{6}{c}{\textit{Dependent variable:}} \\ 
\cline{2-7} 
\\[-1.8ex] & \multicolumn{6}{c}{Tourists flows} \\ 
\\[-1.8ex] & (1) & (2) & (3) & (4) & (5) & (6)\\ 
\hline \\[-1.8ex] 
 Income pc dest. & $-$0.058 & $-$0.033 & $-$0.004 & 0.100 & 0.041 & 0.064 \\ 
  & (0.093) & (0.086) & (0.080) & (0.082) & (0.079) & (0.078) \\ 
  & & & & & & \\ [-0.6em]
 Population dest. & 0.177 & 0.696$^{***}$ & 0.803$^{***}$ & 0.766$^{***}$ & 0.862$^{***}$ & 0.661$^{***}$ \\ 
  & (0.148) & (0.173) & (0.131) & (0.120) & (0.112) & (0.101) \\ 
  & & & & & & \\ [-0.6em]
 Instrength dest. & 0.038 & 0.063 & $-$0.016 & $-$0.076 & $-$0.014 & 0.069$^{**}$ \\ 
  & (0.081) & (0.075) & (0.065) & (0.087) & (0.048) & (0.030) \\ 
  & & & & & & \\ [-0.6em]
 Outstrength dest. & $-$0.088 & $-$1.797$^{***}$ & $-$1.098$^{**}$ & $-$0.457 & $-$0.956$^{***}$ & $-$0.841$^{***}$ \\ 
  & (0.635) & (0.661) & (0.500) & (0.502) & (0.315) & (0.280) \\ 
  & & & & & & \\ [-0.6em]
 Betweenness dest. & 0.079$^{***}$ & 0.031 & 0.071$^{***}$ & 0.074$^{***}$ & 0.082$^{***}$ & 0.044$^{*}$ \\ 
  & (0.018) & (0.023) & (0.019) & (0.023) & (0.024) & (0.023) \\ 
  & & & & & & \\ [-0.6em]
 Authority dest. & $-$0.189$^{*}$ & $-$0.527$^{***}$ & $-$0.467$^{***}$ & $-$0.319 & $-$0.494$^{***}$ & $-$0.726$^{***}$ \\ 
  & (0.110) & (0.163) & (0.167) & (0.235) & (0.163) & (0.151) \\ 
  & & & & & & \\ [-0.6em]
 Hub dest. & 0.746$^{**}$ & 1.037$^{***}$ & 0.883$^{***}$ & 0.900$^{***}$ & 1.067$^{***}$ & 1.667$^{***}$ \\ 
  & (0.311) & (0.273) & (0.212) & (0.258) & (0.248) & (0.555) \\ 
  & & & & & & \\ [-0.6em]
 Efficiency dest. & 0.018 & 0.031$^{*}$ & 0.032$^{**}$ & 0.026 & 0.059$^{***}$ & 0.004 \\ 
  & (0.017) & (0.018) & (0.016) & (0.017) & (0.017) & (0.020) \\ 
  & & & & & & \\ [-0.6em]
 Mountain clst. dest & 0.995$^{***}$ & 1.088$^{***}$ & 1.083$^{***}$ & 1.000$^{***}$ & 0.729$^{***}$ & 1.379$^{***}$ \\ 
  & (0.157) & (0.161) & (0.157) & (0.161) & (0.158) & (0.157) \\ 
  & & & & & & \\ [-0.6em]
Cultural-Lake clst. dest. & 0.384$^{***}$ & 0.490$^{***}$ & 0.477$^{***}$ & 0.626$^{***}$ & 0.534$^{***}$ & 0.902$^{***}$ \\ 
  & (0.105) & (0.115) & (0.103) & (0.112) & (0.093) & (0.097) \\ 
  & & & & & & \\ [-0.6em]
 Cultural heritage inside & 0.054 & $-$0.005 & $-$0.015 & $-$0.031 & $-$0.044 & $-$0.079 \\ 
  & (0.072) & (0.073) & (0.068) & (0.065) & (0.068) & (0.065) \\ 
  & & & & & & \\ [-0.6em]
Ski routes inside & 0.067 & 0.046 & 0.043 & $-$0.004 & 0.014 & 0.034 \\ 
  & (0.049) & (0.050) & (0.047) & (0.046) & (0.049) & (0.043) \\ 
  & & & & & & \\ [-0.6em]
 Farm-houses inside & $-$0.018 & 0.003 & 0.018 & 0.039 & 0.016 & 0.044 \\ 
  & (0.038) & (0.041) & (0.036) & (0.035) & (0.036) & (0.035) \\ 
  & & & & & & \\ [-0.6em]
 Intermodal nodes inside & $-$0.031 & $-$0.293 & $-$0.034 & $-$0.012 & $-$0.175 & $-$0.066 \\ 
  & (0.176) & (0.188) & (0.169) & (0.159) & (0.164) & (0.162) \\ 
  & & & & & & \\ [-0.6em]
 Methane distributors inside & 0.008 & 0.020 & 0.005 & $-$0.019 & $-$0.008 & $-$0.005 \\ 
  & (0.041) & (0.040) & (0.037) & (0.037) & (0.037) & (0.035) \\ 
  & & & & & & \\ [-0.6em]
 Festivals inside & $-$0.097$^{**}$ & $-$0.059 & $-$0.056 & $-$0.022 & $-$0.025 & $-$0.047 \\ 
  & (0.047) & (0.048) & (0.043) & (0.043) & (0.045) & (0.041) \\ 
  & & & & & & \\ [-0.6em]
Museums inside & 0.056 & 0.078 & 0.042 & 0.022 & 0.087 & 0.069 \\ 
  & (0.076) & (0.079) & (0.070) & (0.068) & (0.070) & (0.067) \\ 
  & & & & & & \\ [-0.6em]
Time distance & $-$0.831$^{***}$ & $-$0.848$^{***}$ & $-$1.009$^{***}$ & $-$0.905$^{***}$ & $-$0.950$^{***}$ & $-$0.823$^{***}$ \\ 
  & (0.074) & (0.077) & (0.069) & (0.069) & (0.070) & (0.067) \\ 
  & & & & & & \\ [-0.6em]
 Constant & 9.643$^{***}$ & 7.509$^{***}$ & 9.406$^{***}$ & 8.327$^{***}$ & 10.308$^{***}$ & 8.258$^{***}$ \\ 
  & (1.662) & (1.673) & (1.516) & (1.524) & (1.536) & (1.526) \\ 
  & & & & & & \\ [-0.6em]
\hline \\[-1.8ex] 
Observations & 1,655 & 1,663 & 1,929 & 1,909 & 1,949 & 2,001 \\ 
R$^{2}$ & 0.340 & 0.321 & 0.351 & 0.366 & 0.365 & 0.366 \\ 
Adjusted R$^{2}$ & 0.329 & 0.309 & 0.341 & 0.357 & 0.356 & 0.357 \\ 
\hline 
\hline \\[-1.8ex] 
\textit{Note:}  & \multicolumn{6}{r}{$^{*}$p$<$0.1; $^{**}$p$<$0.05; $^{***}$p$<$0.01} \\ 
\end{tabular}}
\caption{We show the coefficients of drivers of the gravity model highlighted in Figures \ref{fig: gravity coeff month 1}, \ref{fig: gravity coeff month 2}, \ref{fig: gravity coeff month 3}. Column 1-6 refers to models estimated for the first six months of the year 2022. Part II.} 
  \label{fig: app coeff month grav 2} 
\end{table}

\begin{table}[!htbp] \centering 
 \scalebox{0.80}{
\begin{tabular}{@{\extracolsep{5pt}}lcccccc} 
\\[-1.8ex]\hline 
\hline \\[-1.8ex] 
 & \multicolumn{6}{c}{\textit{Dependent variable:}} \\ 
\cline{2-7} 
\\[-1.8ex] & \multicolumn{6}{c}{Tourists flows} \\ 
\\[-1.8ex] & (7) & (8) & (9) & (10) & (11) & (12)\\ 
\hline \\[-1.8ex] 
 Income pc orig. & $-$0.120 & $-$0.073 & 0.059 & $-$0.019 & $-$0.111 & 0.041 \\ 
  & (0.083) & (0.091) & (0.081) & (0.080) & (0.082) & (0.082) \\ 
  & & & & & & \\  [-0.6em]
 Population orig. & 0.661$^{***}$ & 0.435$^{***}$ & 0.693$^{***}$ & 0.625$^{***}$ & 0.645$^{***}$ & 0.651$^{***}$ \\ 
  & (0.113) & (0.121) & (0.115) & (0.115) & (0.121) & (0.119) \\ 
  & & & & & & \\  [-0.6em]
 Instrength orig. & 0.008 & 0.009 & $-$0.030 & 0.025 & $-$0.023 & 0.016 \\ 
  & (0.027) & (0.029) & (0.027) & (0.028) & (0.029) & (0.028) \\ 
  & & & & & & \\  [-0.6em]
 Outstrength orig. & $-$0.495$^{**}$ & $-$0.067 & $-$0.297 & $-$0.517$^{**}$ & $-$0.409$^{*}$ & $-$0.633$^{***}$ \\ 
  & (0.229) & (0.240) & (0.229) & (0.233) & (0.241) & (0.239) \\ 
  & & & & & & \\  [-0.6em]
 Betweenness orig. & 0.016 & $-$0.024 & $-$0.011 & 0.030 & 0.052 & 0.003 \\ 
  & (0.044) & (0.046) & (0.045) & (0.044) & (0.046) & (0.045) \\ 
  & & & & & & \\  [-0.6em]
 Authority orig. & $-$0.373$^{*}$ & $-$0.485$^{**}$ & $-$0.189 & $-$0.506$^{**}$ & $-$0.083 & $-$0.359$^{*}$ \\ 
  & (0.202) & (0.216) & (0.204) & (0.203) & (0.213) & (0.209) \\ 
  & & & & & & \\  [-0.6em]
 Hub orig. & 2.079$^{***}$ & 1.464$^{***}$ & 2.015$^{***}$ & 2.085$^{***}$ & 2.072$^{***}$ & 2.117$^{***}$ \\ 
  & (0.415) & (0.444) & (0.423) & (0.416) & (0.440) & (0.432) \\ 
  & & & & & & \\  [-0.6em]
 Efficiency orig. & 0.021 & $-$0.013 & 0.062 & 0.024 & $-$0.016 & 0.082 \\ 
  & (0.050) & (0.052) & (0.048) & (0.048) & (0.050) & (0.060) \\ 
  & & & & & & \\  [-0.6em]
 Mountain clst. orig. & $-$0.382$^{**}$ & $-$0.348$^{**}$ & 0.022 & 0.078 & $-$0.019 & $-$0.188 \\ 
  & (0.154) & (0.163) & (0.150) & (0.155) & (0.156) & (0.154) \\ 
  & & & & & & \\  [-0.6em]
 Cultural-Lake clst orig. & $-$0.130 & $-$0.008 & 0.049 & 0.087 & 0.128 & 0.007 \\ 
  & (0.089) & (0.094) & (0.091) & (0.090) & (0.094) & (0.093) \\ 
  & & & & & & \\  [-0.6em]
\hline 
\hline \\[-1.8ex] 
\textit{Note:}  & \multicolumn{6}{r}{$^{*}$p$<$0.1; $^{**}$p$<$0.05; $^{***}$p$<$0.01} \\ 
\end{tabular} }
\caption{We show the coefficients of drivers of the gravity model highlighted in Figures \ref{fig: gravity coeff month 1}, \ref{fig: gravity coeff month 2}, \ref{fig: gravity coeff month 3}. Column 7-12 refers to models estimated for six months in the second half of the year 2022. Part I.} 
  \label{fig: app coeff month grav 1 (7-12) } 
\end{table}

\begin{table}[!htbp] \centering 
 \scalebox{0.80}{
\begin{tabular}{@{\extracolsep{5pt}}lcccccc} 
\\[-1.8ex]\hline 
\hline \\[-1.8ex] 
 & \multicolumn{6}{c}{\textit{Dependent variable:}} \\ 
\cline{2-7} 
\\[-1.8ex] & \multicolumn{6}{c}{Tourists flows} \\ 
\\[-1.8ex] & (7) & (8) & (9) & (10) & (11) & (12)\\ 
\hline \\[-1.8ex] 

 Income pc dest. & $-$0.015 & $-$0.255$^{***}$ & $-$0.017 & 0.106 & 0.141$^{*}$ & 0.106 \\ 
  & (0.080) & (0.083) & (0.080) & (0.079) & (0.084) & (0.082) \\ 
  & & & & & & \\  [-0.6em]
 Population dest. & 0.549$^{***}$ & 0.276$^{***}$ & 0.517$^{***}$ & 0.568$^{***}$ & 0.292$^{**}$ & 0.338$^{***}$ \\ 
  & (0.092) & (0.076) & (0.093) & (0.108) & (0.130) & (0.129) \\ 
  & & & & & & \\  [-0.6em]
 Instrength dest. & 0.114$^{***}$ & 0.030 & 0.093$^{***}$ & 0.108$^{**}$ & 0.090$^{**}$ & $-$0.053 \\ 
  & (0.039) & (0.036) & (0.034) & (0.051) & (0.040) & (0.071) \\ 
  & & & & & & \\  [-0.6em]
 Outstrength dest. & $-$0.822$^{***}$ & 0.239 & $-$1.035$^{***}$ & $-$1.900$^{***}$ & $-$0.627 & $-$0.577 \\ 
  & (0.285) & (0.253) & (0.256) & (0.369) & (0.394) & (0.603) \\ 
  & & & & & & \\  [-0.6em]
 Betweenness dest. & 0.024 & 0.133$^{***}$ & 0.049$^{***}$ & 0.029 & 0.078$^{***}$ & 0.060$^{***}$ \\ 
  & (0.025) & (0.025) & (0.018) & (0.020) & (0.020) & (0.016) \\ 
  & & & & & & \\  [-0.6em]
 Authority dest. & $-$0.807$^{***}$ & $-$0.325$^{***}$ & $-$0.634$^{***}$ & $-$0.315$^{**}$ & $-$0.405$^{**}$ & 0.073 \\ 
  & (0.150) & (0.119) & (0.189) & (0.159) & (0.204) & (0.093) \\ 
  & & & & & & \\  [-0.6em]
 Hub dest. & 0.643$^{***}$ & $-$0.161 & 2.242$^{***}$ & 1.498$^{***}$ & 0.888$^{*}$ & 1.214$^{*}$ \\ 
  & (0.185) & (0.243) & (0.562) & (0.260) & (0.477) & (0.684) \\ 
  & & & & & & \\  [-0.6em]
 Efficiency dest. & 0.095$^{***}$ & 0.058$^{***}$ & 0.065$^{***}$ & 0.122$^{***}$ & 0.064$^{***}$ & 0.070$^{***}$ \\ 
  & (0.020) & (0.022) & (0.022) & (0.017) & (0.016) & (0.017) \\ 
  & & & & & & \\  [-0.6em]
 Mountain clst. dest. & 1.203$^{***}$ & 1.084$^{***}$ & 0.732$^{***}$ & 0.454$^{***}$ & 0.493$^{***}$ & 1.124$^{***}$ \\ 
  & (0.142) & (0.148) & (0.158) & (0.158) & (0.170) & (0.143) \\ 
  & & & & & & \\  [-0.6em]
 Cultural-Lake clst. dest. & 0.668$^{***}$ & 0.701$^{***}$ & 0.498$^{***}$ & 0.385$^{***}$ & 0.289$^{***}$ & 0.305$^{***}$ \\ 
  & (0.100) & (0.101) & (0.095) & (0.097) & (0.105) & (0.106) \\ 
  & & & & & & \\  [-0.6em]
 Cultural heritage inside & $-$0.048 & $-$0.018 & 0.017 & $-$0.105 & 0.090 & 0.121$^{*}$ \\ 
  & (0.063) & (0.068) & (0.067) & (0.066) & (0.074) & (0.066) \\ 
  & & & & & & \\  [-0.6em]
 Ski routes inside & $-$0.081$^{*}$ & 0.042 & 0.015 & 0.023 & 0.067 & 0.115$^{**}$ \\ 
  & (0.043) & (0.044) & (0.047) & (0.045) & (0.051) & (0.045) \\ 
  & & & & & & \\  [-0.6em]
 Farm-houses inside & 0.061$^{*}$ & 0.029 & 0.029 & 0.037 & $-$0.007 & $-$0.022 \\ 
  & (0.035) & (0.035) & (0.036) & (0.036) & (0.039) & (0.036) \\ 
  & & & & & & \\  [-0.6em]
 Terminal nodes inside & $-$0.032 & $-$0.046 & $-$0.293$^{*}$ & $-$0.404$^{**}$ & $-$0.328$^{*}$ & $-$0.073 \\ 
  & (0.162) & (0.171) & (0.161) & (0.167) & (0.176) & (0.160) \\ 
  & & & & & & \\  [-0.6em]
 Methane distributors inside & $-$0.053 & 0.003 & 0.041 & 0.049 & 0.091$^{**}$ & 0.086$^{**}$ \\ 
  & (0.035) & (0.038) & (0.037) & (0.037) & (0.040) & (0.037) \\ 
  & & & & & & \\  [-0.6em]
 Festivals inside & 0.016 & $-$0.084$^{*}$ & $-$0.060 & $-$0.013 & $-$0.105$^{**}$ & $-$0.098$^{**}$ \\ 
  & (0.042) & (0.045) & (0.043) & (0.042) & (0.048) & (0.042) \\ 
  & & & & & & \\  [-0.6em]
  Museums inside & 0.012 & 0.049 & $-$0.020 & 0.071 & $-$0.061 & $-$0.109 \\ 
  & (0.065) & (0.070) & (0.070) & (0.068) & (0.077) & (0.071) \\ 
  & & & & & & \\  [-0.6em]
 Time distance inside & $-$0.864$^{***}$ & $-$0.757$^{***}$ & $-$0.890$^{***}$ & $-$0.967$^{***}$ & $-$0.864$^{***}$ & $-$0.875$^{***}$ \\ 
  & (0.067) & (0.070) & (0.068) & (0.069) & (0.073) & (0.073) \\ 
  & & & & & & \\  [-0.6em]
 Constant & 11.601$^{***}$ & 9.319$^{***}$ & 11.689$^{***}$ & 12.330$^{***}$ & 9.299$^{***}$ & 12.202$^{***}$ \\ 
  & (1.504) & (1.584) & (1.532) & (1.506) & (1.538) & (1.519) \\ 
  & & & & & & \\  [-0.6em]
\hline \\[-1.8ex] 
Observations & 1,986 & 1,743 & 1,972 & 1,967 & 1,894 & 1,985 \\ 
R$^{2}$ & 0.377 & 0.349 & 0.359 & 0.373 & 0.341 & 0.334 \\ 
Adjusted R$^{2}$ & 0.368 & 0.338 & 0.350 & 0.364 & 0.331 & 0.324 \\ 
\hline 
\hline \\[-1.8ex] 
\textit{Note:}  & \multicolumn{6}{r}{$^{*}$p$<$0.1; $^{**}$p$<$0.05; $^{***}$p$<$0.01} \\ 
\end{tabular} }
\caption{We show the coefficients of drivers of the gravity model highlighted in Figures \ref{fig: gravity coeff month 1}, \ref{fig: gravity coeff month 2}, \ref{fig: gravity coeff month 3}. Column 7-12 refers to models estimated for six months in the second half of the year 2022. Part II.} 
  \label{fig: app coeff month grav 2 (7-12) } 
\end{table}

\clearpage

\end{document}